\documentclass[manuscript,screen,nonacm]{acmart}

\authorsaddresses{}
\renewcommand\footnotetextcopyrightpermission[1]{}

\usepackage{booktabs}
\usepackage{graphicx}
\usepackage{siunitx}
\graphicspath{{figures/}}

\newcommand{\bR}{\beta_{\mathrm{R}}}
\newcommand{\bAC}{\beta_{\mathrm{AC}}}
\newcommand{\bT}{\beta_{\mathrm{T}}}
\newcommand{\dAC}{\delta_{\mathrm{AC}}}
\newcommand{\dAR}{\delta_{\mathrm{AR}}}
\newcommand{\pperm}{\ensuremath{p_{\mathrm{perm}}}}

\begin{document}

\title{Bias at the Borderline: Who Gets the Benefit of the Doubt in Peer Review? Evidence from ICLR}

\author{Hazem Ibrahim}
\authornote{Corresponding author: hazem.ibrahim@nyu.edu}
\affiliation{%
  \institution{Computer Science, New York University Abu Dhabi}
  \city{Abu Dhabi}
  \country{UAE}
}
\email{hazem.ibrahim@nyu.edu}

\author{Talal Rahwan}
\affiliation{%
  \institution{Computer Science, New York University Abu Dhabi}
  \city{Abu Dhabi}
  \country{UAE}
}

\author{Yasir Zaki}
\affiliation{%
  \institution{Computer Science, New York University Abu Dhabi}
  \city{Abu Dhabi}
  \country{UAE}
}

\begin{abstract}
We study peer review at ICLR, a large machine-learning conference whose complete review record, including rejected submissions, is public. Review there is a two-stage decision: reviewers score each submission, and for the borderline band of papers whose scores do not settle an outcome definitively, an area chair makes a discretionary accept-or-reject call. We ask whether that call is even-handed: do authors from prestigious institutions, from WEIRD (Western, educated, industrialized, rich, and democratic) countries, or from all-male author teams get the benefit of the doubt at the margin? Across ICLR 2019--2025 (31,711 submissions; 10,416 in the borderline band), borderline papers without a top-25-institution author are accepted at a 0.5 to 1.6 percentage point lower rate at the same reviewer scores. The gap arises at the discretionary decision stage, reappears out-of-sample in the pre-registered ICLR 2026 cohort, and concentrates almost entirely among submissions identifiable through a pre-decision arXiv preprint ($-3.4$ vs.\ $-0.2$ points). Equal scores need not mean equal papers, however: an area chair may respond to quality the scores miss, so a decision-rate gap alone cannot distinguish bias from information. We therefore apply a robust outcome test, which concludes discrimination only when the group accepted at a lower rate also realizes better downstream outcomes. We measure five outcomes (citations, disruption, two forms of novelty, eventual venue) across both sides of the decision, including, to our knowledge, the first ``ones that got away'' test of rejected submissions. Our headline result is a null: across a pre-registered family of 27 tests (axes $\times$ outcomes $\times$ decision sides), no disparity concordant with the decision-rate gap survives correction, and we find no evidence that any group faced a higher bar on the outcomes we measure. That null is not an exoneration. ICLR review is double-blind, but a pre-decision preprint pierces the blind through policy-permitted means, and the acceptance gap lives almost entirely in that porosity, a pattern consistent with area chairs inferring quality from \emph{who wrote the paper}. One interpretation consistent with this joint pattern is the use of revealed institutional prestige as a prior, a form of statistical discrimination that outcome tests may not detect, and a practice that double-blind review exists to prevent.
\end{abstract}



\maketitle

\section{Introduction}
\label{sec:intro}

Peer review at ICLR, the machine-learning conference whose public record this paper studies, is a two-stage decision. Reviewers score each submission, and the scores sort most papers into clear accepts and clear rejects. The remainder, the borderline, goes to an area chair (AC) who must make the call the scores did not. That call is consequential and discretionary; re-review experiments suggest a large fraction of marginal decisions would flip on a second reading~\cite{langford2015,beygelzimer2021}. A clear accept leaves nothing to decide; ``the reviews are mixed'' leaves the decision ambiguous.

This paper asks who gets the benefit of the doubt. Experiments show that reviewers favor famous authors and top institutions when identities are visible~\cite{tomkins2017,huber2022nobel}, and authors routinely level the same charge at the discretionary call; whether it is true under double-blind review is unknown. Among borderline submissions to ICLR from 2019 through 2025, do papers by authors from prestigious institutions, from WEIRD countries,\footnote{Western, educated, industrialized, rich, and democratic.} or from all-male author teams (gender inferred from names; Section~\ref{sec:data}) fare better at the discretionary stage than papers with the same reviewer scores?

The first half of the answer is yes, for prestige. At equal reviewer scores, borderline papers with no author from a top-25 institution are accepted at a 0.5 to 1.6 percentage point lower rate across our two data partitions; both survive false-discovery correction, though a stricter permutation test is cleared only out of sample (Section~\ref{sec:results}). The unconditional gap is wider still; ignoring scores altogether, borderline papers lacking a top-25-institution author are accepted at a 1.7 to 3.2 point lower rate. A stage decomposition splits that raw gap into its two possible sources, the scores reviewers assign and the decision made at those scores, and only the decision component survives correction; the disparity sits in the post-review call, not in the reviews themselves. In some years the raw gap is large: in 2021, 55.7\% of borderline papers with a top-institution author were accepted against 45.2\% of those without (Section~\ref{sec:results}). \emph{Where} the gap lives is itself one of our central findings. The penalty is concentrated almost entirely among submissions whose authors were identifiable through a pre-decision arXiv preprint: $-3.4$ points where identifiable against $-0.2$ where not (estimated outside the confirmatory family). Blinding at ICLR is porous exactly where policy permits it to be, and the disparity lives almost entirely in that porosity, which will matter for how the following result below should be read.

The second half of the answer is the reason this paper exists, because a decision-rate gap at equal scores is not yet discrimination. Reviewer scores are a noisy summary of a submission. An AC who reads the paper, the reviews, and the rebuttal may be responding to legitimate quality signals that the scores miss and that happen to correlate with prestige. Separating those accounts is the classic problem of discrimination measurement. The standard tool, the outcome test, is fragile. If the two groups' quality distributions differ away from the threshold, the outcomes of their accepted papers can differ even when both face the same bar, the \emph{infra-marginality} problem~\cite{knowles2001}.

We address these problems with two design choices. First, the borderline restriction: we analyze the submissions scored closest to the estimated acceptance threshold, where decisions are most nearly marginal and the outcome test is informative. Second, the robust outcome test of Gaebler and Goel~\cite{gaebler2025robust}, which concludes discrimination only when two tests agree directionally: the \emph{benchmark test} (a decision-rate disparity conditional on the legitimate signal) and the \emph{outcome test} (an outcome disparity among decided papers). Put plainly, if ACs hold low-prestige papers to a higher bar, two signatures should follow: the low-prestige papers they accept should realize better outcomes (only stronger papers cleared the stricter cut), and their rejected papers should do better downstream as well (a stricter cut turns away papers a fair bar would have taken). The conjunction is one-directional by design. Agreement is evidence of a higher bar; its absence is not proof of equal treatment. We maintain that boundary throughout.

Unlike policing and lending, the settings that motivated the outcome-test literature, peer review makes both sides of the decision observable. In the context of policing, for instance, an analyst cannot see the contraband carried by drivers who were not searched~\cite{anwar2006alternative,pierson2020large}. Rejected ICLR submissions, by contrast, remain public, and most acquire an independent afterlife in the published record. We exploit this to run what is, to our knowledge, the first \emph{reject-side} outcome test of peer review: at equal scores, do the rejected papers of disadvantaged authors go on to do better? We measure downstream outcomes five ways: forward citations, the consolidation--disruption index~\cite{funk2017,wu2019}, two forms of novelty~\cite{uzzi2013,cohan2020specter}, and, for rejected papers, the prestige tier of the venue where the paper was eventually published, traced through the scholarly record (an arXiv posting alone does not count as republication). These are proxies for quality, not quality itself, which is why the family has five members, a concordance rule, and anchors (Section~\ref{sec:framework}).

The result is a null, and an informative one. Across a pre-registered family of 27 outcome cells (three axes $\times$ two decision sides $\times$ the outcome family), no disparity concordant with the benchmark gap survives false-discovery correction. The share of cells even pointing in the bias direction is 56\%, indistinguishable from coin-flipping (bootstrap $p = 0.13$). On the citation anchor and under our conservative variance rule, the observed intervals cap any higher-bar-direction disparity at about 0.06 standard deviations on the prestige and WEIRD axes (the gender outcome analysis is inconclusive under that rule).

Therefore, our analyses provide no evidence that the prestige gap reflects a higher bar on any outcome we measure. That is the absence of the robust test's sufficient condition for discrimination, not proof that the discretion is benign. What the null leaves open is where the discretion's information comes from, and here the identifiability pattern matters. One reading consistent with the full pattern is that area chairs use \emph{revealed identity} as a prior on quality. This is a form of statistical discrimination that outcome tests may not detect, because an accurate prior equalizes marginal outcomes by construction. It is also a practice that double-blind review is designed to prohibit, whether or not the prior is accurate (Section~\ref{sec:discussion}).

\paragraph{Contributions.} (1)~To our knowledge, the first application of the robust outcome test to peer review, and the first outcome test of the \emph{rejected} side of a reviewing process. (2)~A pre-registered audit design (five outcomes, both decision sides, 27 cells, a separate FDR family, anchor/corroborating structure, observed exclusion bounds) as a template for auditing discretionary gatekeeping, where reading a decision-rate disparity as bias is invalid. (3)~Substantive findings for machine-learning peer review: a replicated equal-score prestige gap that sits at the post-review decision stage and concentrates almost entirely where a pre-decision preprint made authors identifiable, with no benchmark-concordant outcome evidence of a higher bar. (4)~An honest account of what the audit cannot rule out, with a pre-registered out-of-sample replication that recovers the discretionary-stage prestige gap on the ICLR 2026 cohort.

Our pre-analysis plan was registered on OSF before the confirmatory analysis was run.\footnote{\url{https://osf.io/ndkuq}}

\section{Related work}
\label{sec:related}

\paragraph{Measuring discrimination in discretionary decisions.}
Separating bias from information in a gatekeeper's decisions is an old problem with a standard toolkit, from taste-based and statistical theories of discrimination~\cite{becker1971economics,phelps1972statistical,arrow1973theory} through audit experiments~\cite{bertrand2004emily} to decision-level tests. Benchmark analyses compare decision rates conditional on observables~\cite{pierson2020large} and inherit omitted-variable critiques; outcome tests compare the realized success of decisions across groups~\cite{knowles2001,ayres2002outcome} and inherit the infra-marginality critique~\cite{knowles2001,anwar2006alternative,simoiu2017problem}. One repair targets the margin via quasi-randomly assigned deciders~\cite{arnold2018racial,hull2021marginal,kleinberg2018human}; another characterizes when outcome comparisons identify bias~\cite{canay2024outcometests,bohren2019dynamics}. Gaebler and Goel~\cite{gaebler2025robust} showed that under a monotone-likelihood-ratio condition, at least one of the two tests is directionally valid, so requiring \emph{concordance} between them is robust to the failure of either leg alone. Our setting adds two advantages: the borderline restriction localizes the analysis near the acceptance threshold, and the public record of rejections makes the outcome of the \emph{negative} decision observable, which policing and lending lack.

\paragraph{Bias in peer review.}
A substantial literature documents disparities in peer review, surveyed in~\cite{lee2013bias,bornmann2011scientific,shah2022challenges}. Experiments that manipulate blinding find reviewers favor famous authors and top institutions when identities are visible~\cite{blank1991double,peters1982peer,ross2006effect,tomkins2017,okike2016singleblind,huber2022nobel}, and double-blind review shifts outcomes for under-represented groups~\cite{budden2008doubleblind}. Grant-review audits find racial and gender gaps whose locus is the evaluation of the applicant rather than the science~\cite{ginther2011race,witteman2019gender}, against a backdrop of low inter-reviewer agreement~\cite{pier2018low} and network effects on judgment~\cite{teplitskiy2018sociology}. Card et al.~\cite{card2020} found that female-authored economics papers accumulate about a quarter more citations than observably similar male-authored papers conditional on referee assessments, evidence of a higher bar; this is outcome-test logic on the accepted side of journal review and the closest precedent for our design. A larger audit of 145 journals, by contrast, finds no comparable gender penalty in manuscript handling~\cite{squazzoni2021peer}. Lepp and Smith~\cite{lepp2025} document language-related disparities in peer review at scale; disadvantage in review is not confined to the axes we code. For machine-learning conferences, observational analyses of OpenReview report score and acceptance disparities by institution rank and author demographics~\cite{tran2020,manzoor2021uncovering}, and experiments isolate resubmission and citation biases~\cite{stelmakh2021prior,stelmakh2023citeseeing}. Preprint posting during double-blind review measurably reveals identity~\cite{bharadhwaj2020,rastogi2022arxiv}, and recent work documents AI-mediated distortions of review~\cite{russolatona2025lottery}. Two features distinguish our study: prior observational work stops at the disparity, without testing whether it reflects a higher bar, and no prior study observes what became of the papers the process turned away. The NeurIPS consistency experiments~\cite{langford2015,beygelzimer2021} established that marginal decisions are noisy; that noise makes room for favoritism, and it is why the borderline is where an audit should look.

\paragraph{Quantifying scientific quality.}
Any outcome test needs an observable stand-in for quality, and every available measure is contestable~\cite{fortunato2018science}. Citations measure attention as much as quality~\cite{bornmann2008citation,aksnes2019citations} and inherit the Matthew effect~\cite{merton1968matthew}; we adjust the citation outcome for prior author-level reach on the prestige axis and treat no single measure as sufficient. The other measures come from the science-of-science literature: the consolidation--disruption index~\cite{funk2017,wu2019,park2023papers}, atypical reference combinations~\cite{uzzi2013}, and embedding novelty built on SPECTER~\cite{cohan2020specter}. Novelty measures carry their own evaluation bias, since novel work is discounted by both reviewers and early citations~\cite{boudreau2016looking,wang2017bias}, one reason they corroborate rather than anchor. Our pre-registration designates the two most defensible measures (citations; eventual venue of rejected papers) as anchors that alone can drive a verdict. To our knowledge, no prior peer-review audit has run a multi-dimensional outcome family under a single multiplicity-corrected decision rule.

\section{Setting and data}
\label{sec:data}

\subsection{Corpus}
We analyze all submissions to the International Conference on Learning Representations (ICLR) for review years 2019 through 2025, drawn from the public OpenReview record: 31,711 submissions with reviews, per-criterion scores, decisions, and discussion threads. ICLR suits this question because its full review record, including the identities and text of \emph{rejected} submissions, becomes public after decisions. Venues that hide rejections cannot support a reject-side outcome test and were excluded after a coverage scan.\footnote{NeurIPS, for example, publishes only accepted submissions' records.} The analyzable unit is a submission with at least two numeric reviewer ratings and a resolved accept/reject decision; withdrawn and desk-rejected submissions are excluded.

\subsection{The borderline band}
For each review year we normalize reviewer ratings within year, average them per paper into a score $\bar{S}$, and estimate the score threshold $\tau$ at which the fitted probability of acceptance equals one half. The \emph{borderline band} is the set of submissions with $|\bar{S}-\tau| \le 0.5$ in within-year standard-deviation units, a half-width fixed in the discovery stage and carried unchanged into the confirmatory analysis. The sample waterfall: 31,711 submissions with posted reviews; 27,142 meet the inclusion rule; 10,416 fall in the band (4,922 accepted, 5,494 rejected), with per-year band counts growing from 395 (2019) to 3,613 (2025).\footnote{The pre-analysis plan's Table~1 projected 13,376 borderline submissions, an estimate made before the final data build (per-year totals scaled by an earlier run's pooled borderline fraction). The realized count under the identical half-width and inclusion rules is 10,416; no band-eligible submission was excluded after band construction.} The band is centered on the fitted 50\%-acceptance threshold (pooled acceptance 47.3\%), but acceptance probability still varies meaningfully across it, from roughly 12\% at the lower edge to 86\% at the upper (Figure~\ref{fig:band}B): the restriction localizes the analysis near the threshold; it does not make every banded decision a coin flip.

\subsection{Group axes}
Each axis codes the group hypothesized to be disadvantaged as $G=1$, fixed in advance. All axes measure \emph{affiliation or name} facts, not authors' identities: ``non-WEIRD'' describes an affiliating country, never a person (construction detail: Appendix~\ref{app:groups}).

\textbf{Institutional prestige.} $G=1$ if no author is affiliated with a top-25 computer-science institution (equivalently, aggregation is by the paper's most-prestigious author), from OpenReview profile affiliations mapped through a CS-rankings crosswalk. Top-10 and continuous-rank codings are registered variants (Section~\ref{sec:robustness}). Institutional prestige has been shown to be among the steepest hierarchies in science: a small set of institutions dominates faculty placement~\cite{clauset2015systematic}, and an institution's place in the hierarchy predicts its researchers' productivity and prominence beyond individual factors~\cite{way2019productivity}. If chairs use institutional signals as a shortcut for quality, papers from outside that narrow apex are the ones at risk at the margin, which makes prestige the axis where discretionary bias is most plausible a priori. Because review is double-blind, using institutional signals presupposes that identity leaks into the process at all; Section~\ref{sec:results-deanon} tests that precondition directly.

\textbf{Country (WEIRD).} $G=1$ if the paper's affiliating country is not WEIRD, i.e.\ not Western, educated, industrialized, rich, and democratic~\cite{henrich2010weirdest}, per a pre-registered country list released with the grid. WEIRD is a coarse, normatively loaded partition; a language-based alternative coding is registered (Section~\ref{sec:robustness}).

\textbf{Gender.} $G=1$ if any author is inferred female from name-based inference, so mixed-gender teams are coded $G=1$; first- and last-author codings are unregistered alternatives. The coding therefore captures \emph{perceived} gender ascribed from names, not self-identified gender. Name-to-gender inference carries differential error across name origins~\cite{santamaria2018comparison}; we suppress cells under ten papers; a ground-truth validation with error-corrected estimates is registered, deferred to the out-of-sample replication (Section~\ref{sec:robustness}). No individual-level identity labels are released.

Resolution rates are reported per year and decision side in Appendix~\ref{app:coverage} (Table~\ref{tab:coverage}), never pooled, because identity resolution is itself a selection process that bears on the reject-side test. The prestige axis resolves essentially fully from 2021 onward and not at all in 2019--2020, because the 2019--2020 records identify authors by e-mail address rather than OpenReview profile id, so no profile affiliations exist for those cohorts; prestige estimates are identified by 2021--2025.

\subsection{Downstream scholarly outcomes}
All outcomes are standardized or percentiled within field and publication year and oriented so that higher is conventionally better. They are proxies for the latent quality the decision is presumed to target, not that quality itself, and several are plausibly shaped by prestige through visibility, networks, and resources (Section~\ref{sec:discussion}). The pre-registered family has five members. The two most defensible (Q1, Q5) are \emph{anchors}; the three contested measures (Q2--Q4) are \emph{corroborating only}, so a signal on a contested measure without anchor support is non-robust by rule. All outcomes are right-censored for recent cohorts; windows and censoring detail are in Appendix~\ref{app:censoring}.

\begin{description}
\item[Q1 --- Forward citations (anchor).] Field-by-publication-year citation percentile over a three-year forward window~\cite{kinney2023semanticscholar}, adjusted on the prestige axis for prior author-level citation reach so the Matthew effect does not masquerade as quality. Available on both sides; on the reject side via tracing rejected submissions to their eventual published record.
\item[Q2 --- Disruption (corroborating).] The consolidation--disruption index CD$_5$~\cite{funk2017,wu2019} over a five-year forward window. The window is unobservable for recent cohorts and the index requires adequate citation coverage; only 15.0\% of focal papers qualify, so this cell is bounded and corroborating, never standalone.
\item[Q3 --- Novelty, embedding distance (corroborating).] One minus the cosine similarity between the paper's SPECTER embedding~\cite{cohan2020specter} and the centroid of same-field papers from preceding years, hence available for all years.
\item[Q4 --- Novelty, atypical recombination (corroborating).] An atypicality measure in the Uzzi et al.~tradition~\cite{uzzi2013}: reference-venue pairings z-scored against a degree-preserving null, summarized by the atypical tail.
\item[Q5 --- Eventual venue, reject side (anchor).] For a borderline-rejected paper, the prestige tier of the venue where it eventually appeared, with no-republication-observed coded as the lowest tier. This avoids conditioning on a successful trace, which is why Q5 anchors the reject side, but it is right-censored: for recent cohorts, ``not observed by the cutoff'' conflates never with not-yet (Appendix~\ref{app:censoring}). Eventual venue is also socially produced (Section~\ref{sec:discussion}), so its anchor role reflects freedom from trace selection, not clean measurement of quality.
\end{description}

On the reject side, citation-type outcomes (Q1--Q3) require tracing a rejected submission to its later record. We match 71.8\% of traced-year in-band rejects (70.2\% of all rejects over all years), report match rates by axis and year, adjust by inverse-probability weighting~\cite{horvitz1952generalization}, and bound the untraced remainder worst-case (Section~\ref{sec:robustness}).

\section{The audit framework}
\label{sec:framework}

\subsection{Why a decision-rate gap is not an answer}
Let $D \in \{0,1\}$ be the accept decision, $G \in \{0,1\}$ the disadvantaged-group indicator, and $\bar{S}$ the mean normalized reviewer score. The natural audit quantity, the gap in $\Pr(D=1 \mid \bar{S}, G)$, is the \emph{benchmark test}. On its own, it is invalid as a discrimination test: the AC observes the paper, the full reviews, and the rebuttal, so an equal-score gap may reflect legitimate signals correlated with group rather than a group-specific bar. The converse tool, the \emph{outcome test}, compares the observed outcomes $Y$ of decided papers across groups at equal scores; if ACs hold group $G=1$ to a higher bar, the marginal $G=1$ accept is better than the marginal $G=0$ accept. Its classic failure is infra-marginality: averages over accepted papers are not averages over marginal papers~\cite{knowles2001,simoiu2017problem}.

Our design uses both tests where each is strongest. The borderline restriction confines the analysis to the papers scored closest to the acceptance threshold, in the spirit of regression-discontinuity-style local analysis~\cite{lee2010regression}, so the analyzed decisions are the most nearly marginal ones the record contains (acceptance probability still varies across the band; Section~\ref{sec:data}). On top of it we apply the robust test of Gaebler and Goel~\cite{gaebler2025robust}: conclude discrimination against $G=1$ only when the benchmark and outcome tests agree directionally, a rule that is valid whenever a monotone-likelihood-ratio property (MLRP) holds for the conditioning signal (informally, a stronger signal indicates a better paper in the same way for both groups). The guarantee is one-directional: the conjunction is \emph{sufficient} evidence of a higher bar; its failure is insufficient evidence of discrimination, not affirmative evidence of equal thresholds. Every verdict below is stated within that boundary, conditional on MLRP, and with a breakdown value $\rho^\ast$: how much more informative the decision-maker's signal would have to be for one group than the other before the verdict could flip (Appendix~\ref{app:rho}).

Our implementation departs from the theorem's canonical form in three ways (a conditional rather than unadjusted benchmark; downstream proxies rather than the decision-maker's utility; a reject-side extension the theorem does not cover), so we present the design as \emph{inspired by} the robust test rather than a direct instantiation; Appendix~\ref{app:theorem} states each departure and how the design contains it. The reject side therefore serves as corroborating descriptive evidence; no verdict rests on it alone.

\subsection{Estimands}
Three stage-decomposition estimands locate any decision-rate gap, and two outcome-test estimands adjudicate it. All conditioning is on a flexible spline in $\bar{S}$, plus field, log paper length, and year fixed effects. Inference clusters at the venue-year level under the conservative max(cluster, HC1) rule of Section~\ref{sec:robustness}; papers are the unit, and dependence across papers sharing authors is not modeled, a limitation disclosed there.

\textbf{Decision side.} $\bT$ is the total in-band group gap in acceptance (average marginal effect on $\Pr(D=1)$, not conditioned on score); $\bR$ is the gap in the reviewer scores themselves; $\bAC$ is the gap in acceptance \emph{conditional} on scores, the \emph{post-score decision-stage} disparity. A gap that appears in $\bT$ and $\bAC$ but not $\bR$ localizes to the post-review decision. The label is deliberate: conditioning on the mean score removes the reviewer stage's summary signal, not everything the decision-maker saw (score dispersion, reviewer confidence, per-criterion scores, rebuttal text, and AC assignment all remain in the residual), so $\bAC$ is a decision-stage disparity, and is an AC effect in the narrow sense only under stronger assumptions. Scores are the ratings as they stand in the public record, after any rebuttal-period updates; pre-rebuttal ratings were not registered, so the study cannot separate score updating during discussion from the final decision. Negative signs indicate disadvantage for $G=1$.

\textbf{Outcome side.} Among borderline-\emph{accepted} papers, $\dAC$ is the group coefficient in $Y \sim G + \mathrm{spline}(\bar{S}) + \mathrm{field} + \log(\mathrm{length}) + \mathrm{year}$; among borderline-\emph{rejected}, public-author papers, $\dAR$ is the same coefficient on the reject side. The sign convention makes the higher-bar signature positive on both sides: if ACs demand more of $G=1$, the $G=1$ papers they accept are better ($\dAC > 0$) and the $G=1$ papers they reject include good work a fair bar would have taken ($\dAR > 0$). The two sides are separate estimands on separate subpopulations.

\subsection{Decision rule, multiplicity, and the informative null}
The confirmatory family has 27 cells: accept side, three axes $\times$ Q1--Q4; reject side, three axes $\times$ Q1--Q5. The family carries its own Benjamini--Hochberg correction~\cite{benjamini1995}, separate from the discovery-stage family, and every cell is reported. Four pre-committed safeguards prevent the grid from manufacturing a positive: (1)~verdicts are defined by \emph{concordance} with the benchmark leg, not by any cell's significance alone; (2)~directions are pre-committed, so a protective-signed disparity cannot be relabeled a finding; (3)~a sign-aggregate omnibus (the fraction of cells with $\delta > 0$, with a cluster block-bootstrap $p$-value) summarizes systematic tilt in one number; and (4)~only anchor outcomes can drive a verdict.

The per-axis verdict has three tiers. \emph{Strong}: benchmark-concordant anchor disparities on both decision sides, each surviving correction. \emph{Moderate}: a concordant anchor on one side, corroborated by a contested measure. \emph{None}: no concordant anchor disparity and a sign-aggregate near its null, read as a well-powered null only when sensitivity analysis says the design could have detected an effect. For that purpose we report effective sample sizes, 80\%-power minimum detectable effects from Monte-Carlo simulation of the full estimator, and equivalence-style bounds~\cite{schuirmann1987comparison,lakens2017equivalence}. A null is reported as informative only when it is accompanied by a statement of which disparities the data exclude.

\paragraph{Partitions and registration.} The analysis uses a discovery/confirmatory split of the corpus, an internal replication device rather than a genuinely held-out sample, and we use ``confirmatory'' in that qualified sense throughout. The full outcome expansion was fixed in a pre-analysis plan registered on OSF before it was run, and ICLR 2026 is the pre-registered clean out-of-sample replication (Section~\ref{sec:results-oos}). Appendix~\ref{app:registration} details partition construction and registration status.

\section{Results}
\label{sec:results}

\begin{figure*}[t]
  \centering
  \includegraphics[width=\textwidth]{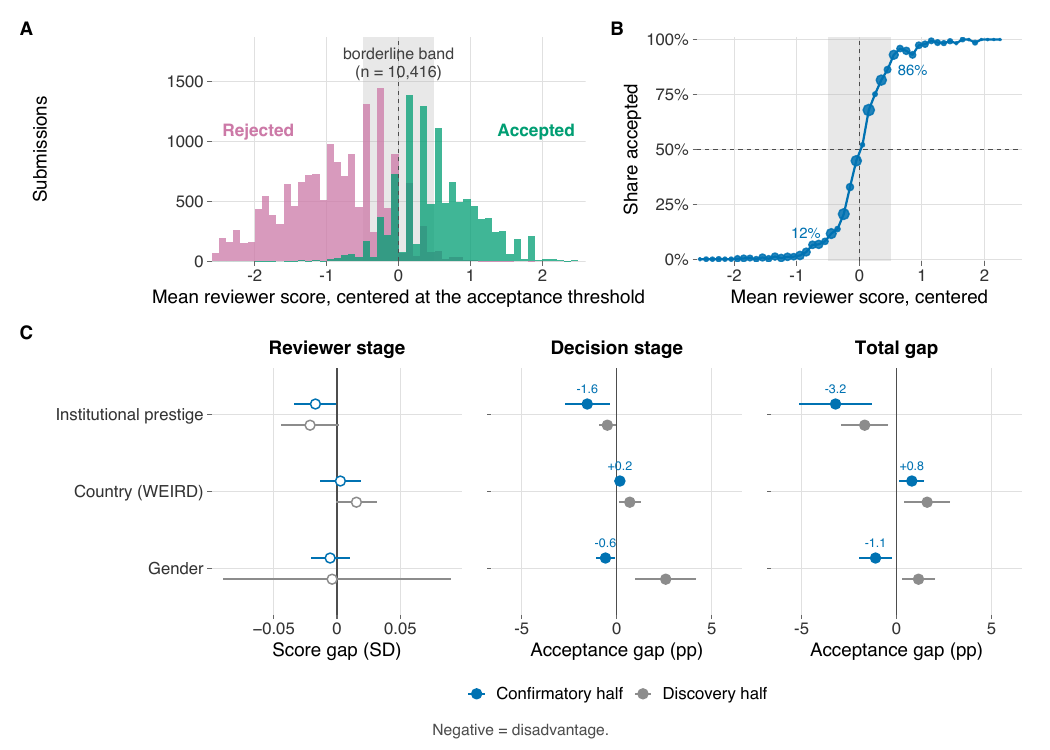}
  \caption{The discretionary regime and the stage decomposition. (A)~Within-year-normalized mean reviewer scores, centered at the fitted 50\%-acceptance threshold ($\bar{S}-\tau$), for the 27{,}142 ICLR submissions meeting the inclusion rule; the shaded band ($|\bar{S}-\tau| \le 0.5$~SD) is the borderline, the 10{,}416 submissions scored closest to the threshold. (B)~Acceptance share across the score axis, rising from 12\% to 86\% across the band. (C)~Stage decomposition of the borderline disparity, discovery and confirmatory halves; the reviewer-score gap ($\bR$, score SD units), the equal-score decision-stage acceptance gap ($\bAC$), and the total in-band acceptance gap ($\bT$), in percentage points. Filled markers survive BH-FDR. The prestige gap is the only robust adverse disparity, and it sits in the decision stage. Whiskers reaching the panel edge are effectively unbounded.}
  \label{fig:band}
\end{figure*}

\subsection{The benchmark: a prestige gap at the discretionary stage}
\label{sec:results-benchmark}

Figure~\ref{fig:band} shows the borderline band (panels A--B) and the stage decomposition of its acceptance disparities (panel~C). Underlying panel~C, the raw within-band acceptance rates, computed directly from the band counts, show the prestige gap year by year: in 2021, 55.7\% of borderline papers with a top-25-institution author were accepted against 45.2\% without; 2024 and 2025 show 49.5\% against 40.2\% and 48.3\% against 44.9\%.

The decomposition places this in the decision conditional on mean reviewer scores, not in the scores themselves. The total in-band acceptance gap for low-prestige papers is $\bT = -1.7$ percentage points in the discovery partition ($q = 0.025$) and $-3.2$ points in the confirmatory partition ($q = 0.009$). The reviewer-score gap $\bR$ is small and does not survive correction in either half ($-0.021$ and $-0.017$ score standard deviations; $q = 0.073$ and $0.056$). The post-score decision-stage gap is $\bAC = -0.5$ points in the discovery half ($p = 0.023$, $q = 0.034$) and $-1.6$ points in the confirmatory half ($p = 0.010$, $q = 0.036$), surviving correction in both.

A supplementary permutation test qualifies how firmly this holds in sample: permuting prestige labels within venue-year ($B = 9{,}999$) and refitting $\bAC$ tests, without cluster asymptotics, the sharp null that the label is unrelated to acceptance through \emph{any} channel. The gap is not separable from chance under it in either half ($\pperm = 0.67$ discovery, $0.20$ confirmatory; all axes: Appendix~\ref{app:spine}), so the max-rule $q$-values overstate in-sample certainty; only on ICLR 2026 does the prestige $\bAC$ clear the permutation test (Section~\ref{sec:results-oos}).

Neither of the other two axes shows an adverse benchmark disparity. The WEIRD-axis estimates are consistently \emph{positive} in both partitions ($\bT$: $+1.6$ points discovery, $+0.8$ confirmatory; $\bAC$: $+0.7$ and $+0.2$, each surviving correction but neither clearing the permutation test), so non-WEIRD authors see no disadvantage at equal scores and, if anything, a small advantage. The gender estimates are unstable rather than directional; the two halves are individually significant with opposite signs ($\bAC$: $+2.6$ points discovery, $-0.6$ confirmatory), and only the discovery half clears the permutation test ($\pperm = 0.03$). We carry both axes through the outcome analysis without a benchmark premise. The benchmark evidence therefore reduces to one finding: an equal-score prestige gap at the post-review decision stage, resting in-sample on the corrected estimates rather than the stricter permutation test, with no stable adverse gap elsewhere.

\subsection{The gap concentrates where authors are identifiable}
\label{sec:results-deanon}

\begin{figure}[t]
  \centering
  \begin{minipage}[c]{0.5\textwidth}
    \includegraphics[width=\linewidth]{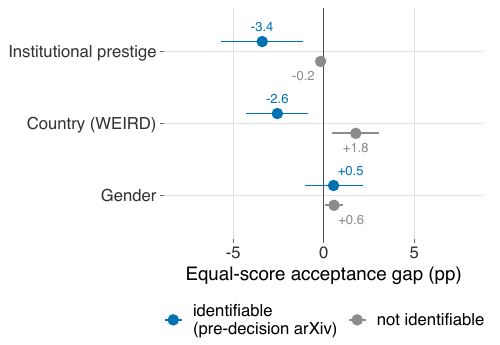}
  \end{minipage}\hfill
  \begin{minipage}[c]{0.44\textwidth}
    \caption{The equal-score gap tracks identifiability. Group acceptance gap at equal reviewer scores among borderline papers whose authors were identifiable through a pre-decision arXiv preprint vs.\ not, per axis; whiskers are 95\% intervals reconstructed from each arm's fit. Interaction $p = 0.005$ (prestige) and $p = 0.002$ (WEIRD); the gender arms are nearly equal in magnitude. Pooled band; suggestive evidence outside the confirmatory family.}
    \label{fig:deanon}
  \end{minipage}
\end{figure}

ICLR review is double-blind on paper, but a submission posted to arXiv before the decision is de-anonymizable in practice~\cite{bharadhwaj2020,rastogi2022arxiv}. Interacting group with pre-decision arXiv availability, the equal-score prestige penalty concentrates among identifiable papers (Figure~\ref{fig:deanon}): $-3.4$ points where a preprint made authors identifiable, $-0.2$ points where it did not (interaction $p = 0.005$). The WEIRD axis shows the same pattern ($-2.6$ against $+1.8$ points, interaction $p = 0.002$); gender shows no penalty in either stratum ($+0.5$ identifiable, $+0.6$ not). This is a pattern \emph{consistent with, but not identifying}, an identity-exposure channel (Section~\ref{sec:discussion}). Three qualifications discipline that reading. First, the pattern requires no chair to violate the norm against looking up submissions: ICLR permits pre-decision preprints, and exposure can be passive, a paper already seen in circulation rather than deliberately searched for. Second, differential posting \emph{rates} alone cannot produce the interaction, since each arm contrasts the two groups within the same stratum; what could mimic it is selection into posting, which is a choice correlated with institution, confidence, seniority, field, and dissemination strategy, and a preprint's existence does not establish that the AC encountered it. Third, the models are pooled-band estimates outside the confirmatory family. What the interaction establishes is that the equal-score gap is a property of papers whose authors were identifiable when the decision was made, not of the band at large.

\subsection{The robust outcome test: no concordant disparity anywhere}
\label{sec:results-outcome}

If the equal-score gap reflects a higher bar rather than information, the accepted low-prestige papers should realize better outcomes at equal scores ($\dAC > 0$) and the rejected pile should contain good papers a fair bar would have admitted ($\dAR > 0$). We test both signatures on all three axes across the five pre-registered outcome dimensions and find neither in benchmark-concordant form; the decision-stage gap leaves no outcome evidence of a higher bar.

\begin{figure*}[t]
  \centering
  \includegraphics[width=\textwidth]{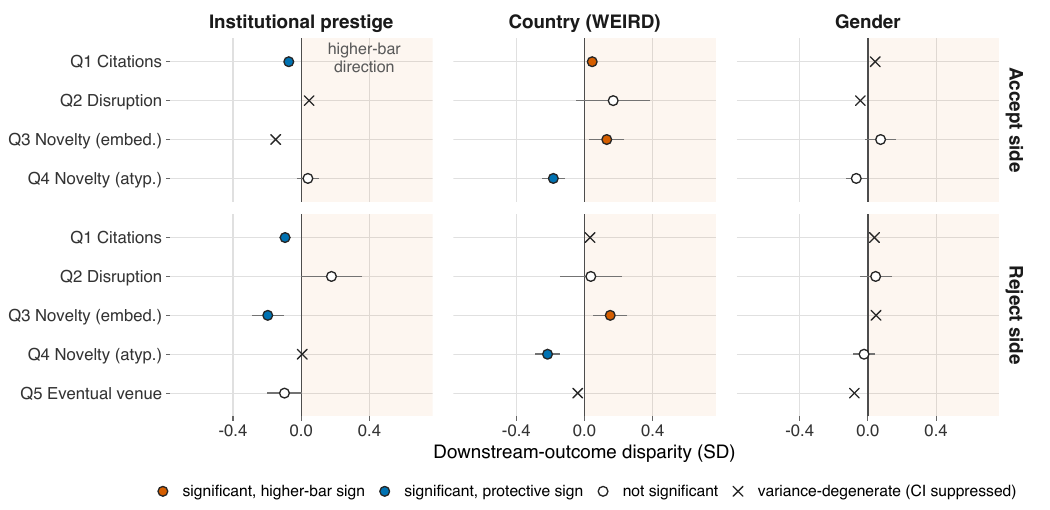}
  \caption{All 27 pre-registered outcome cells: the disparity $\delta$ with 95\% confidence intervals, by axis (columns) and decision side (rows). The shaded half-plane is the higher-bar signature ($\delta > 0$). Filled circles survive BH-FDR (vermillion = bias direction, blue = protective); crosses are variance-degenerate cells, never interpreted. Every prestige survivor is protective-signed; the three bias-direction survivors, all WEIRD, have no benchmark disparity to concord with.}
  \label{fig:grid}
\end{figure*}

Figure~\ref{fig:grid} shows all 27 cells tested; Figure~\ref{fig:concordance} (Appendix~\ref{app:concordance}) pairs the anchor cells with the benchmark leg they would need. Eight cells reject their null after BH correction; their pattern is the opposite of a higher-bar story. Every surviving prestige cell is \emph{negatively} signed: accepted low-prestige borderline papers realize lower citation percentiles than equal-score high-prestige accepts ($\dAC = -0.073$, $q < 10^{-17}$), and rejected ones realize lower citations ($\dAR = -0.095$, $q < 10^{-6}$) and lower embedding novelty ($\dAR = -0.20$, $q < 10^{-3}$). This does not satisfy the robust test's sufficient condition for a higher bar; it points the other way. Nor do the negative disparities certify benign discretion: they are what prestige-linked advantages in visibility and dissemination would produce among equally good papers (Section~\ref{sec:discussion}).

The WEIRD axis shows why the robust test demands both legs. Its accept-side citation anchor is positively signed, the higher-bar signature, and strongly significant ($\dAC = +0.045$, $q < 10^{-6}$); a plain outcome test would declare discrimination here. The benchmark leg it would need is missing, however: the equal-score acceptance estimates for non-WEIRD authors point the \emph{opposite}, protective way in both partitions (pooled $\bAC = +0.4$ points). The corroborating novelty cells split against themselves (embedding novelty $+0.13$ and $+0.15$; atypicality $-0.18$ and $-0.22$): measure disagreement, not a quality story. The gender axis contributes no survivor at all: its citation anchors are positively signed on both sides ($\dAC = +0.043$; $\dAR = +0.038$) but variance-degenerate under the conservative standard-error rule (no finite standard error; Section~\ref{sec:robustness}), and the benchmark leg they would need is sign-unstable across partitions. Ten cells in all, among them the WEIRD reject-side citation anchor and both gender citation anchors, return degenerate variance and stay in the family with near-unit $p$-values, flagged, never interpreted (Appendix~\ref{app:degenerate}).

\begin{table}[t]
  \centering
  \caption{The robust outcome test, summarized. $\bAC$ is the pooled equal-score acceptance gap in percentage points (negative = disadvantage); anchor disparities $\delta$ carry conservative standard errors; ``degen.''\ marks variance-degenerate cells; the survivor column counts anchor cells only. No higher-bar-signed survivor is matched by an adverse benchmark disparity on its axis, which is why every verdict is \textsc{none}. $^{*}q<.05$, $^{**}q<.01$, $^{***}q<.001$.}
  \label{tab:verdicts}
  \small
  \begin{tabular}{lcccccl}
\toprule
Axis & $\beta_{AC}$ (pp) & $\delta_{AC}$ cite & $\delta_{AR}$ cite & $\delta_{AR}$ venue & Higher-bar surv.\ (anchors) & Verdict \\
\midrule
Institutional prestige & $-0.9$ & -0.073 (0.008)$^{***}$ & -0.095 (0.017)$^{***}$ & -0.098 (0.053) & 0 & \textsc{none} \\
Country (WEIRD) & $+0.4$ & 0.045 (0.008)$^{***}$ & degen. & degen. & 1 & \textsc{none} \\
Gender & $+1.0$ & degen. & degen. & degen. & 0 & \textsc{none} \\
\bottomrule
\end{tabular}

\end{table}

The sign aggregate, which resists cherry-picking, tells the same story: 55.6\% of the 27 cells point in the bias direction at all, against a null of 50\% (cluster block bootstrap $p = 0.13$). The pre-registered verdict is \emph{None} on all three axes: for prestige because no defensible outcome disparity concords with the benchmark gap, and for WEIRD and gender because there is no robust benchmark disparity to concord with (Table~\ref{tab:verdicts}). The expansion was designed to overturn the discovery-stage citation-only verdict if citations were hiding bias on another dimension, and it did not: the robust test's condition for concluding discrimination is met on no axis, no outcome, and no decision side.

\subsection{How informative the null is, and on what}
\label{sec:results-power}

Two kinds of statement qualify the null. The first describes the design's sensitivity. Effective sample sizes are roughly 3,600--3,900 per axis on the accept side, and simulation puts the minimum detectable accept-side disparity at 80\% power at about 0.15 standard deviations (Figure~\ref{fig:power}, Appendix~\ref{app:power}). The design could have detected a moderate disparity had one existed; what the data actually rule out is a separate statement. The second comes from the observed intervals and is only as defensible as the variance rule beneath them. Under the conservative rule, the one-sided 95\% bound on a higher-bar-direction citation disparity is $-0.06$~SD for prestige (excluding \emph{any} positive disparity) and $+0.06$~SD for WEIRD, so disparities beyond roughly 0.06~SD are incompatible with the estimates. The gender anchor is variance-degenerate under the same rule, so its outcome analysis is inconclusive; the MDE describes what the design could have detected, not what the gender data exclude. For WEIRD the bound is a magnitude cap rather than equivalence to zero, since its anchor estimate is positive and significant. The corroborating cells are less informative (the WEIRD embedding-novelty bound reaches $+0.22$~SD), one reason they corroborate only. For the acceptance-rate estimand, the achievable 90\% confidence half-width is $\pm 3.8$--$4.0$ percentage points across axes, a precision statement from null simulation rather than an observed equivalence test. The pre-registered per-cell TOST equivalence bounds are reported in Appendix~\ref{app:degenerate}.
Worst-case selection bounds address trace selection directly. Reject-side citation outcomes require tracing rejected papers (70.2\% traced overall), so we compute Lee--Manski bounds~\cite{manski1990,lee2009} treating untraced papers at the extremes. All bracket zero (prestige reject-side citations: $[-0.42, +0.30]$); the point estimates are informative under missing-at-random weighting, and the identified set does not exclude no-effect. Taken together, the null is informative rather than underpowered, concentrated on the prestige and WEIRD citation anchors; gender stays inconclusive.

\subsection{Out-of-sample: the benchmark gap on ICLR 2026}
\label{sec:results-oos}

The pre-registered out-of-sample corpus is ICLR 2026, a cohort never collected or examined before the expanded plan was registered. Its identities and decisions are public for accepted and rejected submissions alike, but downstream outcomes will not accrue for years, so 2026 tests only the benchmark side. We applied the frozen band definition, group codings, and decomposition to its 5{,}486 borderline submissions (Table~\ref{tab:oos}, Appendix~\ref{app:oos}). The prestige pattern reappears. The decision-stage gap is $\bAC = -2.8$ points, and it survives its permutation test ($\pperm = 0.033$, $B = 9{,}999$), which the pooled 2019--2025 estimate did not ($\pperm = 0.27$). With a single venue-year cluster, sandwich $p$-values are not defensible on this cohort, so the permutation test carries all 2026 inference. One caution: affiliation is observed, not assigned, so the permutation test checks whether the gap could arise from a chance arrangement of group labels, not whether affiliation causes the decision. The WEIRD axis, a small acceptance advantage in the main sample, shows a small adverse 2026 gap ($\bAC = -1.5$ points) that its permutation test does not support ($\pperm = 0.23$), and gender is null on every leg. The reappearance of the gap on a cohort collected only after registration, clearing the permutation test, is the strongest single evidence that the gap is real; whether it again carries no higher bar awaits the full replication.

\section{Robustness}
\label{sec:robustness}

\paragraph{Post-rejection revision.}
The sharpest threat to the reject-side test is that rejected papers are revised before they reappear, so the eventual record measures the revision, not the paper the AC saw. We therefore measure revision extent directly: text similarity between the as-rejected and published versions, time to republication, and venue transition (Appendix~\ref{app:revision}). Adding these controls barely moves the reject-side estimates. The prestige citation cell shifts from $-0.095$ to $-0.092$, and revision extent itself does not differ by prestige ($-0.026$, $p = 0.09$). One related difference is genuine. Rejected low-prestige papers land in lower-tier venues than their revision extent predicts ($-0.137$ tiers among the 1{,}149 republished rejects with both versions, $p = 0.007$), which we read as inequality after rejection (Section~\ref{sec:discussion}).

\paragraph{Baseline-quality controls.}
A base-rate account holds that lower-prestige groups produce lower-quality work, so equal scores would mean different expected quality. The pre-registered baseline controls (a leave-one-out author-quality mean, plus institution and country baselines) leave every cell's qualitative pattern unchanged (Appendix~\ref{app:robustnotes}).

\paragraph{Traceability, selection, and right-censoring.}
Tracing rejected papers to their later record succeeds more often for some groups than others (70.0\% vs.\ 60.3\% by prestige in 2025), so the reject-side citation estimates carry inverse-probability weights and worst-case bounds (Section~\ref{sec:results-power}). The eventual-venue anchor (Q5) does not depend on a successful trace, but it is right-censored. For 2024--2025 rejects, ``no publication observed by the cutoff'' usually means ``not published yet,'' so a group that republishes faster would bias exactly these cells. The plan committed two checks in advance and neither changes any verdict (Appendix~\ref{app:censoring}). The first re-estimates Q5 within a fixed 34-month window, the longest available to every 2019--2023 cohort, so no group gains from extra elapsed time; it shows no higher-bar signal (eight of nine estimates negative, disadvantaged rejects landing in \emph{lower} tiers; none clears $0.05$ under the max rule; four cells degenerate). The second check, a hazard model of time to republication, confirms the concern: non-WEIRD rejects do republish faster (hazard ratio $1.6$ pooled, $p = 0.003$; gender $1.7$ confirmatory, $p = 0.02$; prestige indistinguishable), which is precisely the differential the fixed window removes. Censoring bias is thus live in the as-run Q5 cells, but no verdict rests on them; the fixed-window estimates, immune to that differential, show no higher-bar signal.

\paragraph{Few-cluster inference and degenerate cells.}
Our inference has only seven venue-year clusters, too few for the usual clustered standard errors to be reliable (early runs returned missing or implausibly small cluster variances). Every cell therefore reports the larger of the clustered and heteroskedasticity-robust (HC1) variances. This max rule is conservative only in an informal sense, with no coverage guarantee, so we lean on two further checks. The permutation test of Section~\ref{sec:results-benchmark} is the finite-sample one, and it is stringent. No in-sample $\bAC$ clears it, so the max-rule $q$-values overstate the certainty of the prestige cell, and only the ICLR-2026 gap survives; the larger total gap $\bT$ does survive correction in both halves, with the same caveat. The second check is a CR2 (Bell--McCaffrey--Satterthwaite) adjustment on all 27 outcome cells and it changes no verdict. The accept-side anchors stay significant, the exclusion bounds move little, and cell-level significance shifts in both directions at ${\sim}3$ degrees of freedom, including one newly significant gender cell that lacks the benchmark leg a verdict would need (Appendix~\ref{app:degenerate}). Finally, the ten degenerate cells (Section~\ref{sec:results-outcome}) arise where inverse-probability weights meet thin coverage-filtered samples; no verdict rests on one (Appendix~\ref{app:degenerate}).

In addition, the three axes correlate and each estimate averages over the other two; intersectional cells (e.g., prestige $\times$ country) were not pre-registered and are too thin at the borderline, so an intersection-confined disparity could escape every reported cell.

\paragraph{Robustness variants run and registered.}
Any single specification embeds choices: band width, prestige coding, citation window, controls. The pre-analysis plan therefore registered a grid of variants over these choices (baseline-quality and revision controls, inverse-probability weighting with Lee--Manski bounds, the max(cluster, HC1) variance rule, the band grid, four prestige codings, the language coding, two- and five-year citation windows), and every variant that evaluates the full robust test returns \textsc{none} on all three axes.

The band grid behaves as a threshold-local gap should. At wider bands ($\pm 0.75$ and $\pm 1.0$ SD) the confirmatory equal-score prestige gap attenuates to $-0.5$ and $-0.6$ points (both surviving correction); at the narrowest band ($\pm 0.25$ SD) the estimate is larger ($-2.6$ points) but the halved sample no longer clears correction (Appendix~\ref{app:bandwidth}).

The gap is a boundary effect, separating institutions inside the reputational (CS-rankings) top 25 from everyone else, rather than an elite gradient within the top ranks. Under a top-10 coding the gap nearly vanishes and flips sign between partitions ($+0.3$ discovery, $-0.1$ confirmatory). A three-category coding puts the advantage over outside-top-25 institutions at $+0.5$ points for the top 10 and $+1.7$ points for ranks 11--25 (pooled, both $p < 0.02$). A continuous-rank coding gives ranked institutions a $+1.8$-point advantage at rank 25 (pooled penalized-likelihood~\cite{firth1993} $p = 0.010$, stable across halves), with a within-ranked gradient an order of magnitude smaller. Under the SciSciNet crosswalk, an $h$-index top 25 that includes industrial laboratories, the gap is sign-unstable across halves ($+0.5$ discovery, $-1.1$ confirmatory); the disparity is specific to the reputational coding as well as to its boundary. The band grid, top-10 and continuous-rank codings, language coding, citation windows, and baseline controls are the plan's stability exhibits; the three-category gradient and SciSciNet crosswalk are post hoc. Still unrun: the registered gender error correction, committed to the out-of-sample replication.

\section{Discussion}
\label{sec:discussion}

Peer review's hardest calls are the borderline ones, where scores settle nothing and an area chair must decide. This study asked whether those discretionary calls are even-handed: whether papers from less prestigious institutions, non-WEIRD countries, or with inferred-female authors are held to a higher bar. A gap in acceptance rates alone cannot answer that question, because the area chair sees more than the scores and may be responding to real quality signals. We therefore paired the acceptance-rate comparison with a second question: did the papers of the group accepted less often go on to do better, on either side of the decision? Only when both answers agree does the robust outcome test conclude discrimination.

The answer comes in two parts that have to be held together. First, there is a real gap. Borderline papers without a top-25-institution author are accepted less often at the same reviewer scores, in both halves of our data and again in ICLR 2026, a cohort collected only after the plan was frozen. The gap sits in the post-review decision, not in the scores themselves, and it nearly disappears when authors were not identifiable through a pre-decision arXiv preprint. The second part is the check that most audits never run, and it returns a null with power behind it. If low-prestige papers faced a higher bar, the ones that survived it should look better afterwards, and the rejected pile should hide good work. Across five outcome measures and both sides of the decision, neither signature appears: accepted low-prestige work does not overperform on our proxies, and the reject side shows no sign of systematically undervalued papers, though worst-case bounds do not exclude an undervalued tail (Section~\ref{sec:results-power}). What we can claim is therefore narrow: a prestige disparity in post-score acceptance decisions, with no outcome evidence that a higher bar produced it.

What explains a gap with no higher bar behind it? The strongest clue is where the gap lives. At equal scores, the prestige penalty is $-3.4$ points among papers identifiable through a pre-decision preprint and $-0.2$ among those not (Section~\ref{sec:results-deanon}). Double-blind review is porous exactly where policy permits it, and nearly the whole disparity sits there. Information read from an anonymous manuscript does not care whether its authors posted to arXiv; a prior about \emph{who wrote the paper} does. The pattern is therefore what one would expect if area chairs use revealed identity as a prior on paper quality. None of this requires misconduct: preprints are permitted, encountering one can be passive, and the pattern indicts the porosity of the blind, not chair conduct outside the rules. A higher bar, including one that weights prestige beyond what it actually predicts, would have left the outcome signature we tested for; an \emph{accurate} prestige prior would not, because it equalizes later outcomes by construction. That is a form of statistical discrimination, and it is what double-blind review exists to prevent, whether or not the prior is accurate. We cannot show any area chair actually inferred an institution, nor that such a prior would be accurate, and the interaction sits outside the confirmatory family; whether calibrated use of identity is ever acceptable is a normative question our test cannot settle. Separating paper-derived from identity-derived information would take designs that manipulate identifiability itself, such as enforced anonymous preprinting, and on this evidence a conference's preprint policy is a fairness lever, not a housekeeping rule.

The lesson for fairness auditing generalizes. Discretionary gatekeeping (hiring screens, editorial desks, grant panels, moderation appeals) is routinely audited by decision-rate disparities alone~\cite{pierson2020large,ginther2011race}, an inference invalid in both directions~\cite{canay2024outcometests,kleinberg2018human}. Where the negative decision leaves a public trace, the rejected side is auditable too; the template: restrict to the margin, pre-register outcomes with anchors, demand concordance, report what the null excludes. Institutions that want their gatekeeping trusted should design for that observability. The gender citation estimates are positively signed on both decision sides (${\sim}0.04$~SD) but variance-degenerate, and the benchmark leg is sign-unstable; a benchmark disparity on fresh data with a defensible outcome signature would flip the verdict. Separately, rejected low-prestige papers land in lower-tier venues than their revision extent predicts (Section~\ref{sec:robustness}): not the area chair's bar but inequality \emph{after} rejection, since the capacity to absorb a rejection and resubmit upward is unevenly distributed~\cite{way2019productivity,hofstra2020diversity}.

The verdict comes with limits. The robust test's conjunction is sufficient evidence of a higher bar, not necessary, so every null here reads ``no robust-test evidence,'' never proof of absence. A higher bar on a dimension our proxies miss, or one offset by prestige's advantages in visibility and citations, is consistent with everything we report. The test's validity condition (MLRP) can fail if scores are more informative for one group; our pre-registered sensitivity index puts the breakdown value an order of magnitude above what the data imply ($\rho^\ast = 15.1$ against $\rho = 1.01$; Appendix~\ref{app:rho}), though the index is heuristic. The outcome proxies are contestable and several are themselves shaped by prestige, so the null is not affirmative evidence of well-calibrated discretion. The gender axis relies on name-based inference, which measures \emph{perceived} gender and misclassifies differentially across name origins; every gender statement is conditional on that instrument (Section~\ref{sec:robustness}). Finally, the design is silent about bias upstream of the borderline: reviewer scoring, desk decisions, who submits at all. A fair discretionary margin atop a biased pipeline is still a biased pipeline. The ICLR 2026 replication will rerun the full robust test once outcomes accrue (Section~\ref{sec:results-oos}).

\section{Ethics statement}
\label{sec:ethics}

This is an observational analysis of a review record the venue makes public by design, including for rejected submissions; no interaction with authors, reviewers, or area chairs occurred, and no non-public data was accessed. We release aggregate results only: every cell is suppressed below ten papers, no individual-level identity labels are released or retained, and no attempt is made to re-identify reviewers or area chairs. Author identities on rejected submissions are public on OpenReview, but we treat them as sensitive: nothing in our released artifacts links a named rejected paper to a group coding or an outcome.

Name-based gender inference is an ethically loaded instrument: it misclassifies, differentially by name origin, and it operationalizes gender as a binary many authors do not occupy. We use it because auditing gender fairness at scale requires some group signal; we mitigate by suppressing small cells, confining claims to group aggregates, and labeling every gender result as conditional on the instrument (Section~\ref{sec:robustness}). A ground-truth validation with error-corrected estimates is registered for the out-of-sample replication and has not yet been run; until it is, the gender axis should be read as exploratory. Prestige and country codings likewise carry resolution error, and claims on those axes are conditional on the affiliation instrument.

The findings carry a dual-use risk we state plainly. A null verdict at the discretionary margin could be misread as ``peer review is fair'' and used to dismiss complaints this study does not address. Our result concerns one venue, one stage of the pipeline, and the axes and effect sizes stated; it is compatible with bias upstream in reviewing, in desk decisions, in access, and below our detection thresholds, and the gender verdict is conditional on an uncorrected inference instrument. We have written the claims to resist that misreading and ask that the study be cited with its scope.

\paragraph{Generative AI usage statement.}
We disclose that large language models were used in the production of this paper. OpenAI Codex wrote portions of the data-pipeline code, while Claude Fable 5 was used to format the paper and improve clarity of the writing. All study design, hypotheses, the pre-analysis plan's commitments, and all decisions about claims and their strength are the authors'; every reported number was produced by the pre-registered analysis pipeline and verified against its outputs by the authors, who take full responsibility for the content.

\bibliographystyle{ACM-Reference-Format}
\bibliography{refs}

\appendix
\section{Relation to the robust-test theorem}
\label{app:theorem}

Our implementation departs from the canonical form of the Gaebler--Goel result in three ways, so we present the design as inspired by the robust test rather than a direct instantiation. First, the theorem is stated for an unadjusted decision-rate comparison, whereas our benchmark leg is a regression-adjusted group coefficient conditional on the score spline, field, year, and length. A conditional analogue is natural where the legitimate signal is observed, but we do not claim the averaged coefficient inherits the formal guarantee if signs vary across strata; the omnibus and per-cell reporting are designed so that such heterogeneity is visible rather than absorbed. Second, the theorem's outcome is the utility the decision-maker acts on; ours are downstream proxies for it. The chain we rely on (latent paper quality, the information the AC uses, the acceptance utility, later scholarly outcomes, our five measured proxies) has a link the data cannot certify at each step, which is why the family carries anchors, a concordance rule, and explicit measurement caveats rather than a claim that any proxy \emph{is} quality. Third, the theorem concerns the positive-decision side; the reject-side mirror is our extension, motivated by the same threshold logic applied locally at the margin. We treat it as corroborating descriptive evidence rather than a formally equivalent second leg, and no verdict rests on the reject side alone.

\section{Partitions and registration status}
\label{app:registration}

The discovery/confirmatory partition is a deterministic split of submissions by hashed submission id, fixed in the pipeline before the confirmatory analysis was run and stratified by venue-year by construction. The unit is the paper, so an author can appear in both partitions, and the preprocessing code was developed with access to the full corpus. The partition is therefore an internal replication device, not a genuinely held-out sample, and the main text uses ``confirmatory'' in that qualified sense throughout.

The benchmark result and a single-outcome (citation) accept-side test were completed first as a discovery stage. The full outcome expansion (both sides, the five-outcome family, the concordance rule, and all decision thresholds) was fixed in a pre-analysis plan registered on OSF before the expansion was run, on the same corpus. We label the expansion \emph{pre-specified-but-post-hoc}: the discovery citation cell had been unsealed, the remaining 26 cells had not, and the plan binds the analysis but cannot convert a shared data partition into a held-out one. ICLR 2026 is the pre-registered clean out-of-sample replication; its benchmark side is reported in Section~\ref{sec:results-oos}.

\section{Group-axis construction}
\label{app:groups}

\textbf{Prestige.} Affiliations are the author-entered OpenReview profile affiliations as of the submission year where the profile records history, mapped through a CS-rankings crosswalk. $G=1$ if no author is at a top-25 institution. An institution that resolves but is unranked (including industry labs) counts as not-top-25; an author whose affiliation does not resolve contributes nothing, and the paper is missing on this axis only if no author resolves.

\textbf{Country (WEIRD).} The paper's country is the modal affiliation country of its authors in the bibliometric record where available; otherwise $G=1$ if any author's profile-reported country is non-WEIRD. The pre-registered country list is released with the grid.

\textbf{Gender.} $G=1$ if any author is inferred female; authors whose names cannot be classified contribute nothing, and the paper is missing only if no author's name classifies. Mixed-gender teams are therefore coded $G=1$ (``inferred-female authorship present'').

\section{Outcome windows and censoring}
\label{app:censoring}

Citations (Q1) use a three-year forward window from the Semantic Scholar corpus, with two- and five-year windows reported as robustness variants (Section~\ref{sec:robustness}); where the windowed count is unavailable the total count at the data snapshot is used. Percentiling within publication year compares equally mature papers with one another, but it does not extend any cohort's observation window: for the youngest cohorts the window is truncated at the snapshot and their cells rest on early citations. The disruption index (Q2) requires a five-year forward window, which recent cohorts cannot supply; its coverage filter (15.0\% of focal papers) is skewed toward older cohorts. For eventual venue (Q5), ``no publication observed by the cutoff'' conflates never-republished, not-yet-republished, and match failure, so recent-cohort cells compress toward the lowest tier for both groups. The two pre-committed refinements (the plan's fixed-window Q5 primary over the 2019--2023 cohorts, and a republication-hazard model treating recent papers as censored) are discussed in Section~\ref{sec:robustness}; Tables~\ref{tab:q5fw} and~\ref{tab:q5haz} report them in full. The fixed-window primary uses the same estimator and controls as the as-run Q5 cell (group + score spline + field + log-length + venue-year fixed effects, max-rule variance) on the ordinal tier within a common 1{,}020-day (34-month) window from the decision, the longest available to every 2019--2023 cohort at the 2025-11-12 snapshot. The hazard model is a discrete-time logit on monthly person-periods with the same controls plus duration bins. Both are pre-committed robustness checks outside the BH family; no multiplicity correction is applied or needed for their role.

\begin{table}[h]
  \centering
  \caption{Registered fixed-window Q5 primary: eventual-venue tier within a common 34-month window, 2019--2023 reject cohorts. Negative estimates place the disadvantaged group's rejects in lower tiers, the direction opposite a higher bar. $^{\dagger}$~max-rule variance degenerate, suppressed.}
  \label{tab:q5fw}
  \footnotesize
  \begin{tabular}{llrrrrr}
\toprule
Axis & Partition & Est. & SE & $p$ & $n$ & Clusters \\
\midrule
Institutional prestige & pooled & $-0.102$ & 0.053 & 0.053 & 1742 & 3 \\
Institutional prestige & discovery & $-0.135$ & ---$^{\dagger}$ & ---$^{\dagger}$ & 882 & 3 \\
Institutional prestige & confirmatory & $-0.057$ & ---$^{\dagger}$ & ---$^{\dagger}$ & 860 & 3 \\
Country (WEIRD) & pooled & $-0.049$ & ---$^{\dagger}$ & ---$^{\dagger}$ & 1917 & 5 \\
Country (WEIRD) & discovery & $0.032$ & 0.077 & 0.683 & 976 & 5 \\
Country (WEIRD) & confirmatory & $-0.137$ & 0.075 & 0.069 & 941 & 5 \\
Gender & pooled & $-0.081$ & ---$^{\dagger}$ & ---$^{\dagger}$ & 2301 & 5 \\
Gender & discovery & $-0.147$ & 0.080 & 0.066 & 1176 & 5 \\
Gender & confirmatory & $-0.025$ & 0.089 & 0.774 & 1125 & 5 \\
\bottomrule
\end{tabular}

\end{table}

\begin{table}[h]
  \centering
  \caption{Pre-committed republication-hazard check: discrete-time logit on monthly person-periods, 2019--2023 reject cohorts. HR $>1$: the disadvantaged group's rejects republish faster. $^{\ddagger}$~cell too thin for the fixed-effects logit to converge.}
  \label{tab:q5haz}
  \footnotesize
  \begin{tabular}{llrrrrr}
\toprule
Axis & Partition & HR & $p$ & Papers & Person-months & Events \\
\midrule
Institutional prestige & pooled & \multicolumn{2}{c}{fit failed$^{\ddagger}$} & 1742 & 54920 & 164 \\
Institutional prestige & discovery & 0.77 & 0.268 & 882 & 27611 & 91 \\
Institutional prestige & confirmatory & 1.48 & 0.104 & 860 & 26961 & 73 \\
Country (WEIRD) & pooled & 1.59 & 0.003 & 1917 & 59746 & 202 \\
Country (WEIRD) & discovery & \multicolumn{2}{c}{fit failed$^{\ddagger}$} & 976 & 30241 & 110 \\
Country (WEIRD) & confirmatory & 1.69 & 0.043 & 941 & 29162 & 92 \\
Gender & pooled & 1.13 & 0.397 & 2301 & 72390 & 218 \\
Gender & discovery & \multicolumn{2}{c}{fit failed$^{\ddagger}$} & 1176 & 36850 & 118 \\
Gender & confirmatory & 1.68 & 0.020 & 1125 & 35192 & 100 \\
\bottomrule
\end{tabular}

\end{table}

\section{Full pre-registered grid}
\label{app:grid}

Table~\ref{tab:grid} reports every cell of the pre-registered 27-cell outcome family: the outcome disparity $\hat\delta$ with its 95\% confidence interval, raw and BH-adjusted $p$-values, sample size, and an effective-sample-size diagnostic. The diagnostic is a variance-ratio measure, $n_{\mathrm{eff}} = n \cdot \widehat{\mathrm{se}}^2_{\mathrm{naive}} / \widehat{\mathrm{se}}^2_{\mathrm{full}}$, comparing the precision of the full specification (spline, fixed effects, weights) to an unadjusted group comparison; it is not a Kish weighting ESS, and values modestly above $n$ indicate that the covariates absorb residual outcome variance (a precision gain) rather than an error. Values far above $n$ indicate weight-driven variance pathology and feed the degeneracy flag of Appendix~\ref{app:degenerate}. Cells flagged $^{\dagger}$ are variance degenerate under that criterion; their confidence intervals and effective sample sizes are suppressed rather than reported at implausible magnitudes.

\begin{table}[h]
  \centering
  \caption{All 27 pre-registered outcome cells. $^{\dagger}$~variance degenerate (flagged, never interpreted).}
  \label{tab:grid}
  \footnotesize
  \begin{tabular}{lllcccccc}
\toprule
Axis & Side & Outcome & $\hat\delta$ & 95\% CI & $p$ & $q$ & $n$ & $n_{\mathrm{eff}}$ \\
\midrule
Institutional prestige & Accept & Q1 Citations & $-0.073$ & $[-0.09, -0.06]$ & $<\!10^{-18}$ & $<\!10^{-17}$ & 4,323 & 4,280 \\
Institutional prestige & Accept & Q2 Disruption & $0.046$ & --- & 1.000 & 1.000 &   612 & ---$^{\dagger}$ \\
Institutional prestige & Accept & Q3 Novelty (embed.) & $-0.150$ & --- & 1.000 & 1.000 & 1,684 & ---$^{\dagger}$ \\
Institutional prestige & Accept & Q4 Novelty (atyp.) & $0.039$ & $[-0.03, 0.11]$ & 0.248 & 0.479 & 3,713 & 3,737 \\
Institutional prestige & Reject & Q1 Citations & $-0.095$ & $[-0.13, -0.06]$ & $<\!10^{-7}$ & $<\!10^{-6}$ & 3,289 & 1,072 \\
Institutional prestige & Reject & Q2 Disruption & $0.177$ & $[-0.00, 0.36]$ & 0.054 & 0.145 &   342 &   187 \\
Institutional prestige & Reject & Q3 Novelty (embed.) & $-0.196$ & $[-0.29, -0.10]$ & $<\!10^{-4}$ & 0.000 & 1,890 & 1,866 \\
Institutional prestige & Reject & Q4 Novelty (atyp.) & $0.006$ & --- & 1.000 & 1.000 & 3,035 & ---$^{\dagger}$ \\
Institutional prestige & Reject & Q5 Eventual venue & $-0.098$ & $[-0.20, 0.01]$ & 0.063 & 0.156 & 1,742 & 2,201 \\
Country (WEIRD) & Accept & Q1 Citations & $0.045$ & $[0.03, 0.06]$ & $<\!10^{-7}$ & $<\!10^{-6}$ & 4,677 & 4,497 \\
Country (WEIRD) & Accept & Q2 Disruption & $0.168$ & $[-0.05, 0.39]$ & 0.130 & 0.269 &   871 &   864 \\
Country (WEIRD) & Accept & Q3 Novelty (embed.) & $0.130$ & $[0.03, 0.23]$ & 0.014 & 0.046 & 1,870 & 1,665 \\
Country (WEIRD) & Accept & Q4 Novelty (atyp.) & $-0.183$ & $[-0.25, -0.11]$ & $<\!10^{-6}$ & $<\!10^{-5}$ & 4,055 & 3,514 \\
Country (WEIRD) & Reject & Q1 Citations & $0.032$ & --- & 0.999 & 1.000 & 3,464 & ---$^{\dagger}$ \\
Country (WEIRD) & Reject & Q2 Disruption & $0.037$ & $[-0.15, 0.22]$ & 0.693 & 1.000 &   477 &   192 \\
Country (WEIRD) & Reject & Q3 Novelty (embed.) & $0.151$ & $[0.05, 0.25]$ & 0.003 & 0.012 & 2,034 & 1,834 \\
Country (WEIRD) & Reject & Q4 Novelty (atyp.) & $-0.217$ & $[-0.29, -0.14]$ & $<\!10^{-8}$ & $<\!10^{-7}$ & 3,209 & 3,024 \\
Country (WEIRD) & Reject & Q5 Eventual venue & $-0.040$ & --- & 1.000 & 1.000 & 1,917 & ---$^{\dagger}$ \\
Gender & Accept & Q1 Citations & $0.043$ & --- & 0.999 & 1.000 & 4,863 & ---$^{\dagger}$ \\
Gender & Accept & Q2 Disruption & $-0.044$ & --- & 1.000 & 1.000 &   979 & ---$^{\dagger}$ \\
Gender & Accept & Q3 Novelty (embed.) & $0.075$ & $[-0.02, 0.17]$ & 0.108 & 0.243 & 1,957 & 1,876 \\
Gender & Accept & Q4 Novelty (atyp.) & $-0.068$ & $[-0.13, -0.01]$ & 0.028 & 0.085 & 4,234 & 4,164 \\
Gender & Reject & Q1 Citations & $0.038$ & --- & 0.997 & 1.000 & 3,840 & ---$^{\dagger}$ \\
Gender & Reject & Q2 Disruption & $0.046$ & $[-0.05, 0.14]$ & 0.345 & 0.621 &   573 &   521 \\
Gender & Reject & Q3 Novelty (embed.) & $0.048$ & --- & 1.000 & 1.000 & 2,267 & ---$^{\dagger}$ \\
Gender & Reject & Q4 Novelty (atyp.) & $-0.022$ & $[-0.09, 0.04]$ & 0.500 & 0.845 & 3,455 & 3,413 \\
Gender & Reject & Q5 Eventual venue & $-0.079$ & --- & 1.000 & 1.000 & 2,301 & ---$^{\dagger}$ \\
\bottomrule
\end{tabular}

\end{table}

\section{Benchmark and spine estimates}
\label{app:spine}

Table~\ref{tab:spine} reports the stage-decomposition spine on the discovery and confirmatory partitions: the reviewer-score gap $\beta_R$ (score SD units), the equal-score acceptance gap $\beta_{AC}$, and the total in-band acceptance gap $\beta_T$ (both in acceptance-probability units), per axis, with BH-adjusted $q$-values within each partition's family.

\begin{table}[h]
  \centering
  \caption{Stage-decomposition spine, discovery and confirmatory halves.}
  \label{tab:spine}
  \small
  \begin{tabular}{llcccccc}
\toprule
 & & \multicolumn{3}{c}{Discovery} & \multicolumn{3}{c}{Confirmatory} \\
\cmidrule(lr){3-5} \cmidrule(lr){6-8}
Estimand & Axis & Est. & $p$ & $q$ & Est. & $p$ & $q$ \\
\midrule
$\beta_R$ & Institutional prestige & $-0.021$ & 0.065 & 0.073 & $-0.017$ & 0.044 & 0.056 \\
$\beta_R$ & Country (WEIRD) & $0.015$ & 0.064 & 0.073 & $0.003$ & 0.749 & 0.749 \\
$\beta_R$ & Gender & $-0.004$ & 1.000 & 1.000 & $-0.005$ & 0.489 & 0.550 \\
$\beta_{AC}$ & Institutional prestige & $-0.005$ & 0.023 & 0.034 & $-0.016$ & 0.010 & 0.036 \\
$\beta_{AC}$ & Country (WEIRD) & $0.007$ & 0.021 & 0.034 & $0.002$ & 0.025 & 0.038 \\
$\beta_{AC}$ & Gender & $0.026$ & 0.002 & 0.016 & $-0.006$ & 0.022 & 0.038 \\
$\beta_T$ & Institutional prestige & $-0.017$ & 0.008 & 0.025 & $-0.032$ & 0.001 & 0.009 \\
$\beta_T$ & Country (WEIRD) & $0.016$ & 0.009 & 0.025 & $0.008$ & 0.017 & 0.038 \\
$\beta_T$ & Gender & $0.012$ & 0.011 & 0.025 & $-0.011$ & 0.012 & 0.036 \\
\bottomrule
\end{tabular}

\end{table}

The supplementary permutation test for $\beta_{AC}$ (group permuted within venue-year, $B = 9{,}999$; no seven-cluster asymptotics) gives $p_{\mathrm{perm}}$ of 0.67 (discovery) and 0.20 (confirmatory) for prestige (0.27 pooled), 0.57 and 0.90 for WEIRD, and 0.03 and 0.61 for gender. Only the gender discovery cell is individually significant under it, and it reverses sign in the confirmatory half (Section~\ref{sec:results-benchmark}).

\section{Band-width dependence of the equal-score gap}
\label{app:bandwidth}

\begin{figure}[h]
  \centering
  \includegraphics[width=\columnwidth]{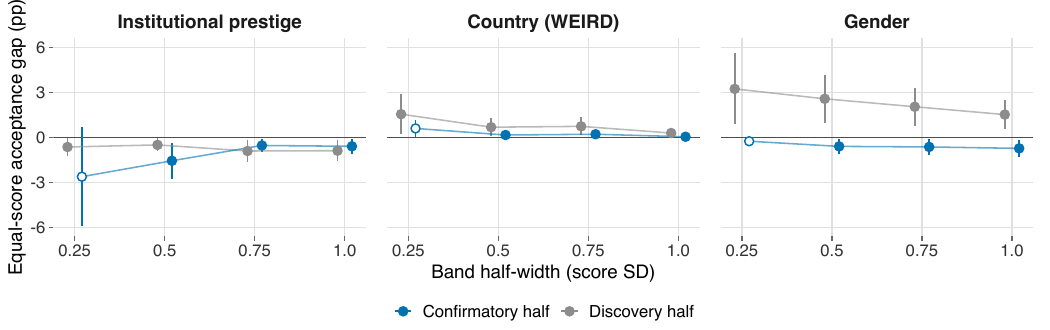}
  \caption{The equal-score acceptance gap across band half-widths, per axis; whiskers are 95\% intervals, filled markers survive BH-FDR within their run.}
  \label{fig:bandwidth}
\end{figure}

Figure~\ref{fig:bandwidth} traces the equal-score acceptance gap $\beta_{AC}$ across the pre-registered band half-widths (0.25, 0.5, 0.75, 1.0 score SD), per axis and partition; the estimates are reported in Section~\ref{sec:robustness}. The attenuation with band width is the direction of dependence a threshold-local effect would produce, though the narrow-band estimate is too imprecise to establish it on its own. Every band returns the robust-test verdict \textsc{none} on all three axes.

\section{Coverage, match rates, and selection}
\label{app:coverage}

Table~\ref{tab:coverage} reports group-resolution rates per year and decision side for each axis: the share of borderline submissions whose group coding resolves. As noted in Section~\ref{sec:data}, the prestige axis resolves essentially fully from 2021 onward and not at all in 2019--2020; WEIRD coverage is partial in 2019--2020 and near-full thereafter.

\begin{table}[h]
  \centering
  \caption{Group-resolution rates (share of borderline submissions with a resolved group coding), by year, decision side, and axis.}
  \label{tab:coverage}
  \small
  \begin{tabular}{lcccccc}
\toprule
 & \multicolumn{2}{c}{Prestige} & \multicolumn{2}{c}{WEIRD} & \multicolumn{2}{c}{Gender} \\
\cmidrule(lr){2-3} \cmidrule(lr){4-5} \cmidrule(lr){6-7}
Year & Accept & Reject & Accept & Reject & Accept & Reject \\
\midrule
2019 & 0.00 & 0.00 & 0.62 & 0.23 & 0.99 & 0.99 \\
2020 & 0.00 & 0.00 & 0.76 & 0.42 & 1.00 & 1.00 \\
2021 & 1.00 & 1.00 & 0.98 & 0.98 & 1.00 & 1.00 \\
2022 & 1.00 & 1.00 & 0.98 & 0.98 & 1.00 & 1.00 \\
2023 & 1.00 & 1.00 & 0.99 & 0.99 & 1.00 & 1.00 \\
2024 & 1.00 & 1.00 & 0.99 & 0.99 & 1.00 & 1.00 \\
2025 & 1.00 & 1.00 & 1.00 & 1.00 & 1.00 & 1.00 \\
\bottomrule
\end{tabular}

\end{table}

Reject-side citation-type outcomes require tracing a rejected submission to its eventual published record. Table~\ref{tab:match} reports trace rates by axis, year, and group; the differential motivates the inverse-probability weights carried by every reject-side citation estimate and the worst-case selection bounds of Table~\ref{tab:bounds}, which treat all untraced papers at the outcome extremes (Lee--Manski). All bounds bracket zero (Section~\ref{sec:results-power}).

\begin{table}[h]
  \centering
  \caption{Reject-side trace rates by axis, year, and group ($G{=}1$ = disadvantaged group).}
  \label{tab:match}
  \small
  \begin{tabular}{lcccccc}
\toprule
 & \multicolumn{2}{c}{Prestige} & \multicolumn{2}{c}{WEIRD} & \multicolumn{2}{c}{Gender} \\
\cmidrule(lr){2-3} \cmidrule(lr){4-5} \cmidrule(lr){6-7}
Year & $G{=}0$ & $G{=}1$ & $G{=}0$ & $G{=}1$ & $G{=}0$ & $G{=}1$ \\
\midrule
2019 & --- & --- & 1.00 & 1.00 & 0.98 & 0.97 \\
2020 & --- & --- & 1.00 & 1.00 & 0.99 & 0.99 \\
2021 & 0.78 & 0.71 & 0.81 & 0.62 & 0.76 & 0.74 \\
2022 & 0.70 & 0.68 & 0.69 & 0.71 & 0.70 & 0.68 \\
2023 & 0.64 & 0.59 & 0.65 & 0.56 & 0.64 & 0.60 \\
2024 & 0.72 & 0.65 & 0.70 & 0.67 & 0.68 & 0.69 \\
2025 & 0.70 & 0.60 & 0.68 & 0.62 & 0.66 & 0.65 \\
\bottomrule
\end{tabular}

\end{table}

\begin{table}[h]
  \centering
  \caption{Worst-case (Lee--Manski) selection bounds for reject-side outcomes.}
  \label{tab:bounds}
  \small
  \begin{tabular}{llcc}
\toprule
Axis & Reject-side outcome & Worst-case bounds & $n$ \\
\midrule
Institutional prestige & Citations (Q1) & $[-0.42, 0.30]$ & 5,494 \\
Institutional prestige & Disruption (Q2) & $[-8.70, 9.04]$ & 5,494 \\
Institutional prestige & Novelty, embed.\ (Q3) & $[-3.07, 2.93]$ & 5,494 \\
Country (WEIRD) & Citations (Q1) & $[-0.33, 0.37]$ & 5,494 \\
Country (WEIRD) & Disruption (Q2) & $[-8.20, 8.75]$ & 5,494 \\
Country (WEIRD) & Novelty, embed.\ (Q3) & $[-2.82, 3.02]$ & 5,494 \\
Gender & Citations (Q1) & $[-0.30, 0.35]$ & 5,494 \\
Gender & Disruption (Q2) & $[-7.80, 8.04]$ & 5,494 \\
Gender & Novelty, embed.\ (Q3) & $[-2.64, 2.72]$ & 5,494 \\
\bottomrule
\end{tabular}

\end{table}

\section{Power, MDE, and equivalence}
\label{app:power}

\begin{figure}[h]
  \centering
  \begin{minipage}[c]{0.5\textwidth}
    \includegraphics[width=\linewidth]{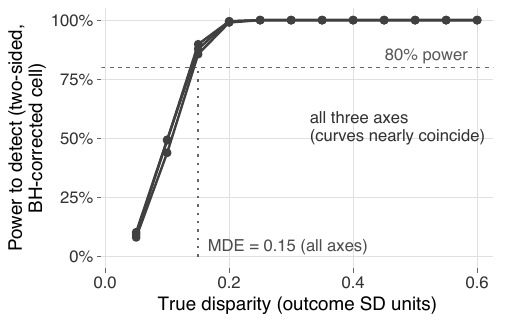}
  \end{minipage}\hfill
  \begin{minipage}[c]{0.44\textwidth}
    \caption{Design sensitivity. Monte-Carlo power curves (1{,}000 replications per point) for the accept-side outcome disparity $\dAC$; the 80\%-power minimum detectable effect is ${\sim}0.15$ outcome SD on each axis. Bounds on what the data exclude come from the observed intervals in Section~\ref{sec:results-power}.}
    \label{fig:power}
  \end{minipage}
\end{figure}

Table~\ref{tab:power} reports, per axis: the borderline-band sample, the effective sample size after the score spline and fixed effects absorb within-band variance (via the variance inflation factor of the group indicator), the 80\%-power minimum detectable effects for the benchmark ($\beta_{AC}$, log-odds) and outcome ($\delta_{AC}$, outcome SD) estimands from Monte-Carlo simulation (1{,}000 replications per grid point), and the achievable 90\% confidence half-width on the equal-score acceptance gap in average-marginal-effect percentage points (the median CI half-width under null simulation at the realized effective sample size). All of these are design-sensitivity quantities: they say what the design could detect or bound, not what the data have excluded. The observed exclusion bounds quoted in Section~\ref{sec:results-power} are one-sided 95\% limits from the estimated cells of Table~\ref{tab:grid}; the registered per-cell TOST equivalence bounds, the two-sided counterpart, are reported in Table~\ref{tab:cr2}. Power curves are shown in Figure~\ref{fig:power}.

\begin{table}[h]
  \centering
  \caption{Effective sample sizes, minimum detectable effects, and equivalence bounds.}
  \label{tab:power}
  \small
  \begin{tabular}{lcccccc}
\toprule
Axis & $n_{\mathrm{band}}$ & $n_{\mathrm{eff}}$ & VIF & MDE $\beta_{AC}$ & MDE $\delta_{AC}$ (SD) & CI half-width (pp) \\
\midrule
Institutional prestige & 4,517 & 3,699 & 1.22 & $0.15$ & $0.15$ & $\pm 3.9$ \\
Country (WEIRD) & 4,517 & 3,672 & 1.23 & $0.15$ & $0.15$ & $\pm 4.0$ \\
Gender & 4,517 & 3,835 & 1.18 & $0.15$ & $0.15$ & $\pm 3.8$ \\
\bottomrule
\end{tabular}

\end{table}

\section{Degenerate cells}
\label{app:degenerate}

Section~\ref{sec:robustness} motivates the max(cluster, HC1) variance rule. A cell is flagged \emph{variance degenerate} when its reported standard error is missing, exceeds ten (an order of magnitude above any estimate in the family), or its IPW effective sample size exceeds four times the raw count (weight concentration). Ten of the 27 cells are flagged (Table~\ref{tab:grid}); they concentrate where inverse-probability weights meet thin coverage-filtered samples: the CD$_5$ disruption index is defined for only 15.0\% of focal papers (CD$_3$: 31.4\%). Flagged cells remain in the BH family with their honest near-unit $p$-values; no verdict rests on any of them.

The CR2 (Bell--McCaffrey) adjustment with model-based Satterthwaite degrees of freedom, run on all 27 cells (Section~\ref{sec:robustness}), is computed from a rank-truncated design (one exactly collinear column, from the field and venue-year factors, is dropped per cell) and returns finite intervals everywhere, including the flagged cells. It changes no verdict. The accept-side citation anchors stay individually significant (prestige $p = 0.0007$, WEIRD $p = 0.019$; Satterthwaite dof ${\approx}3$), and the one-sided bounds of Section~\ref{sec:results-power} move little (prestige $-0.06$~SD unchanged; WEIRD $+0.07$~SD, from $+0.06$). At three degrees of freedom cell-level significance moves in both directions: some max-rule survivors weaken (the WEIRD reject-side citation anchor, $p = 0.13$ from $p = 0.02$), and among the flagged cells one becomes individually significant, the gender accept-side citation anchor ($+0.043$~SD, $p = 0.017$, uncorrected; one-sided bound $+0.07$~SD). Without a directionally stable benchmark leg the latter supports no robust verdict; we read it as the same positively signed, one-legged pattern discussed in Section~\ref{sec:discussion}. Table~\ref{tab:cr2} reports the full CR2 grid. The registered BH family is defined on the max-rule $p$-values; recomputing BH on the CR2 $p$-values ($q_{\mathrm{CR2}}$, Table~\ref{tab:cr2}) yields four survivors (prestige accept-Q1 and reject-Q3, WEIRD accept-Q4 and reject-Q4), no benchmark-concordant pair among them, and the gender anchor does not survive it: every verdict is unchanged under either correction. The table's final column reports the pre-registered per-cell TOST equivalence bound, the smallest realized-quality disparity the cell rules out at the 0.05 level under the max rule; it is the two-sided (magnitude) counterpart of the directional one-sided bounds of Section~\ref{sec:results-power}.

\begin{table}[h]
  \centering
  \caption{CR2 (Bell--McCaffrey--Satterthwaite) supplementary inference and pre-registered TOST equivalence bounds, all 27 outcome cells. Est.: group coefficient (outcome SD); dof: Satterthwaite degrees of freedom; $q_{\mathrm{CR2}}$: BH within the 27-cell family on CR2 $p$-values (supplementary; the registered family uses the max-rule $p$-values); 1-sided bd.: one-sided 95\% CR2 bound in the higher-bar direction; TOST bd.: smallest magnitude excluded at TOST-0.05 under the max rule ($^{\dagger}$~max-rule variance degenerate, suppressed).}
  \label{tab:cr2}
  \footnotesize
  \begin{tabular}{llcrrrrrrr}
\toprule
Axis & Side & Cell & Est. & SE$_{\mathrm{CR2}}$ & dof & $p_{\mathrm{CR2}}$ & $q_{\mathrm{CR2}}$ & 1-sided bd. & TOST bd. \\
\midrule
Institutional prestige & Accept & Q1 & $-0.073$ & 0.005 & 2.9 & 0.001 & 0.009 & $-0.06$ & 0.09 \\
Institutional prestige & Accept & Q2 & $0.046$ & 0.052 & 1.1 & 0.531 & 0.573 & $0.34$ & ---$^{\dagger}$ \\
Institutional prestige & Accept & Q3 & $-0.150$ & 0.058 & 2.7 & 0.090 & 0.173 & $-0.01$ & 0.24 \\
Institutional prestige & Accept & Q4 & $0.039$ & 0.022 & 3.2 & 0.169 & 0.268 & $0.09$ & 0.09 \\
Institutional prestige & Reject & Q1 & $-0.095$ & 0.019 & 2.9 & 0.016 & 0.063 & $-0.05$ & 0.12 \\
Institutional prestige & Reject & Q2 & $0.177$ & 0.009 & 1.1 & 0.024 & 0.072 & $0.22$ & 0.33 \\
Institutional prestige & Reject & Q3 & $-0.196$ & 0.013 & 2.5 & 0.001 & 0.013 & $-0.16$ & 0.28 \\
Institutional prestige & Reject & Q4 & $0.006$ & 0.073 & 3.0 & 0.942 & 0.942 & $0.18$ & 0.07 \\
Institutional prestige & Reject & Q5 & $-0.098$ & 0.029 & 1.9 & 0.085 & 0.173 & $-0.01$ & ---$^{\dagger}$ \\
Country (WEIRD) & Accept & Q1 & $0.045$ & 0.010 & 3.0 & 0.019 & 0.063 & $0.07$ & 0.06 \\
Country (WEIRD) & Accept & Q2 & $0.168$ & 0.089 & 2.1 & 0.191 & 0.287 & $0.42$ & 0.35 \\
Country (WEIRD) & Accept & Q3 & $0.130$ & 0.036 & 2.9 & 0.038 & 0.102 & $0.22$ & 0.22 \\
Country (WEIRD) & Accept & Q4 & $-0.183$ & 0.011 & 3.3 & 0.000 & 0.007 & $-0.16$ & 0.24 \\
Country (WEIRD) & Reject & Q1 & $0.032$ & 0.016 & 2.9 & 0.133 & 0.224 & $0.07$ & 0.06 \\
Country (WEIRD) & Reject & Q2 & $0.037$ & 0.035 & 2.0 & 0.402 & 0.486 & $0.14$ & 0.19 \\
Country (WEIRD) & Reject & Q3 & $0.151$ & 0.064 & 2.6 & 0.113 & 0.204 & $0.31$ & ---$^{\dagger}$ \\
Country (WEIRD) & Reject & Q4 & $-0.217$ & 0.028 & 3.1 & 0.004 & 0.026 & $-0.15$ & 0.28 \\
Country (WEIRD) & Reject & Q5 & $-0.040$ & 0.036 & 2.1 & 0.381 & 0.486 & $0.06$ & ---$^{\dagger}$ \\
Gender & Accept & Q1 & $0.043$ & 0.011 & 3.9 & 0.017 & 0.063 & $0.07$ & ---$^{\dagger}$ \\
Gender & Accept & Q2 & $-0.044$ & 0.044 & 2.3 & 0.414 & 0.486 & $0.07$ & ---$^{\dagger}$ \\
Gender & Accept & Q3 & $0.075$ & 0.015 & 3.6 & 0.011 & 0.057 & $0.11$ & 0.15 \\
Gender & Accept & Q4 & $-0.068$ & 0.025 & 4.2 & 0.049 & 0.119 & $-0.02$ & 0.12 \\
Gender & Reject & Q1 & $0.038$ & 0.017 & 4.0 & 0.082 & 0.173 & $0.07$ & 0.05 \\
Gender & Reject & Q2 & $0.046$ & 0.049 & 2.2 & 0.437 & 0.492 & $0.18$ & 0.12 \\
Gender & Reject & Q3 & $0.048$ & 0.047 & 3.5 & 0.373 & 0.486 & $0.15$ & ---$^{\dagger}$ \\
Gender & Reject & Q4 & $-0.022$ & 0.048 & 4.0 & 0.664 & 0.689 & $0.08$ & 0.10 \\
Gender & Reject & Q5 & $-0.079$ & 0.074 & 3.2 & 0.361 & 0.486 & $0.09$ & ---$^{\dagger}$ \\
\bottomrule
\end{tabular}

\end{table}

\section{The robust test as paired forests}
\label{app:concordance}

\begin{figure}[h]
  \centering
  \includegraphics[width=\textwidth]{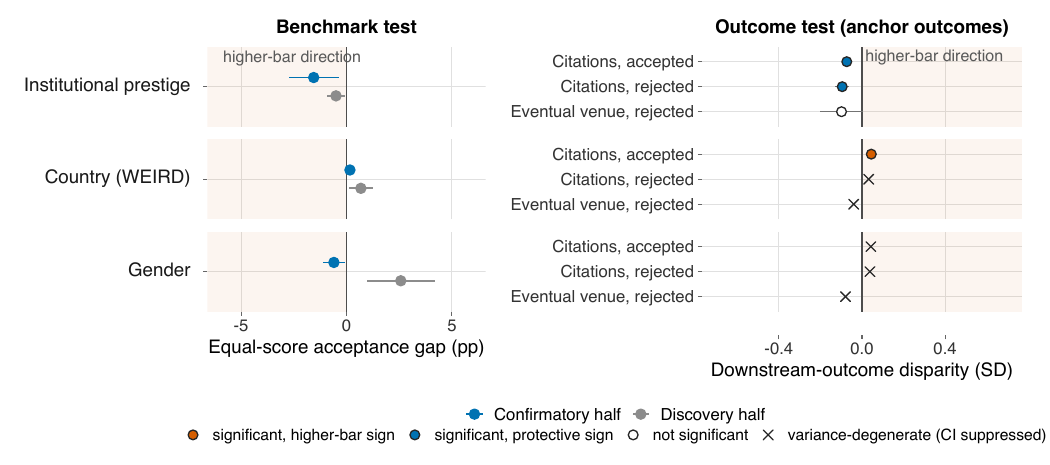}
  \caption{The robust outcome test as paired forests, one row block per group axis. Left: the benchmark test (equal-score acceptance gap $\bAC$, percentage points), discovery and confirmatory halves. Right: the outcome test on the anchor outcomes (disparity $\delta$, SD units). Each column shades its discrimination-consistent direction; the four-way significance coding applies to the outcome panel only (the benchmark panel distinguishes halves). Concluding discrimination requires significant estimates in \emph{both} shaded regions of the same row, and no axis meets the conjunction. Full grid: Figure~\ref{fig:grid}.}
  \label{fig:concordance}
\end{figure}

Figure~\ref{fig:concordance} presents the robust outcome test of Section~\ref{sec:results-outcome} as paired forests, pairing each axis's benchmark leg with its anchor-outcome cells.

\section{Out-of-sample benchmark decomposition: ICLR 2026}
\label{app:oos}

Table~\ref{tab:oos} reports the frozen-rule benchmark decomposition on the ICLR 2026 cohort discussed in Section~\ref{sec:results-oos}.

\begin{table}[h]
  \centering
  \small
  \caption{Out-of-sample benchmark decomposition on ICLR 2026 (5{,}486 borderline submissions, 40.8\% accepted). $\bR$: reviewer-score gap (score SD); $\bAC$: decision-stage acceptance gap; $\bT$: total in-band gap (percentage points); $\pperm$: permutation $p$ for $\bAC$ ($B = 9{,}999$). Negative denotes disadvantage. With a single venue-year cluster, sandwich $p$-values are not defensible on this cohort (they are estimator-sensitive), so the permutation test carries the inference. Downstream outcomes are structurally unavailable for 2026, so the full verdict awaits the replication.}
  \label{tab:oos}
  \begin{tabular}{lcccc}
  \toprule
  Axis & $\bR$ & $\bAC$ & $\pperm$ & $\bT$ \\
  \midrule
  Institutional prestige & $-0.003$ & $-2.8$ & 0.033 & $-3.0$ \\
  Country (WEIRD)        & $-0.010$ & $-1.5$ & 0.235 & $-2.1$ \\
  Gender                 & $+0.001$ & $+1.4$ & 0.299 & $+1.4$ \\
  \bottomrule
  \end{tabular}
\end{table}

\section{Additional robustness notes}
\label{app:robustnotes}

\paragraph{Baseline-quality controls.}
Adding the pre-registered baseline controls (Section~\ref{sec:robustness}) to every expansion cell leaves each cell's qualitative pattern unchanged, with the expected attenuation where the control is nearly the treatment. Per the plan these are robustness exhibits, never primary.

\paragraph{Citation windows and the language coding.}
At two- and five-year citation windows the citation-anchor cells shift at the third decimal (prestige accept-side $-0.074$ and $-0.072$ SD against $-0.073$ at the primary three-year window) and every verdict is unchanged. Five-year windows are truncated for the 2024--2025 cohorts, so that variant leans on older cohorts. The language-based country coding (a TOEFL-exemption partition of the affiliating country) returns a null equal-score gap in both halves and pooled (permutation $p = 0.19$, $0.47$, $0.31$), but its enrichment source resolves a country for only nine percent of the borderline band, so we read it as a limited-coverage check, not a powered replication of the WEIRD null.

\section{MLRP sensitivity: breakdown values}
\label{app:rho}

The robust-test verdict is conditional on a monotone-likelihood-ratio property for the conditioning signal (Section~\ref{sec:framework}), which fails if the signal is differentially informative across groups. We quantify sensitivity with a one-parameter variance-ratio violation family: the data-implied violation $\rho$ is the ratio of within-band reviewer-score variance between groups, $\mathrm{Var}(\bar{S} \mid G{=}1)/\mathrm{Var}(\bar{S} \mid G{=}0)$, and the breakdown value $\rho^\ast = 1 + |\delta_{AC}| / (k\,(|\beta_{AC}| + \epsilon))$ with $k = 0.5$, $\epsilon = 10^{-3}$ is the violation magnitude at which the cell's verdict could flip. This is a stylized, pre-committed sensitivity index scaled by the observed effect magnitudes, not a quantity derived from the theorem itself; we report it so the MLRP conditionality is auditable rather than rhetorical. On the pooled-band citation anchor, the data-implied $\rho$ is close to one on every axis while $\rho^\ast$ is an order of magnitude larger (prestige: $\rho = 1.01$ against $\rho^\ast = 15.1$; WEIRD: $0.91$ against $18.7$; gender: $0.99$ against $9.4$), so no verdict is within range of the observed differential informativeness. No cell triggers the pre-registered degrade-to-invalid rule ($\rho > \rho^\ast$).

\section{Revision-extent measurement}
\label{app:revision}

For every borderline-rejected paper matched to an eventual published version, we measure revision extent as (i)~the SPECTER-embedding text similarity between the as-rejected OpenReview version and the eventual version, (ii)~days to republication, and (iii)~the venue-tier transition. Table~\ref{tab:revision} Panel~A re-estimates every reject-side cell adding these controls; Panel~B regresses each revision-extent measure on group within the matched set. Revision behavior does differ by group in Panel~B: rejected low-prestige and non-WEIRD papers land in lower venue tiers ($-0.137$, $p = 0.007$; $-0.295$, $p < 10^{-7}$) and non-WEIRD revisions depart further from the rejected text ($-0.043$, $p = 0.002$). What defends the reject-side test is Panel~A: no reject-side estimate moves materially once these revision-extent controls are added (the prestige citation cell shifts from $-0.095$ to $-0.092$), so differential post-rejection improvement does not explain the reject-side estimates, and the venue-tier differential is read as a post-rejection inequality in its own right (Section~\ref{sec:discussion}).

\begin{table}[h]
  \centering
  \caption{Revision-extent robustness for the reject-side test.}
  \label{tab:revision}
  \small
  \begin{tabular}{llccc}
\toprule
\multicolumn{5}{l}{\emph{Panel A: reject-side cells with revision-extent controls}} \\
Axis & Outcome & Primary $\hat\delta_{AR}$ (SE) & With controls (SE) & $n$ \\
\midrule
Institutional prestige & Q1 Citations & $-0.095$ (0.017) & $-0.092$ (0.017) & 3,289 \\
Institutional prestige & Q2 Disruption & $0.177$ (0.092) & $0.165$ (0.086) &   342 \\
Institutional prestige & Q3 Novelty (embed.) & $-0.196$ (0.049) & $-0.209$ (0.048) & 1,890 \\
Institutional prestige & Q4 Novelty (atyp.) & degen. & $0.013$ (0.066) & 3,035 \\
Institutional prestige & Q5 Eventual venue & $-0.098$ (0.053) & degen. & 1,742 \\
Country (WEIRD) & Q1 Citations & degen. & $0.038$ (0.014) & 3,464 \\
Country (WEIRD) & Q2 Disruption & $0.037$ (0.093) & $0.032$ (0.097) &   477 \\
Country (WEIRD) & Q3 Novelty (embed.) & $0.151$ (0.051) & $0.126$ (0.054) & 2,034 \\
Country (WEIRD) & Q4 Novelty (atyp.) & $-0.217$ (0.037) & $-0.208$ (0.037) & 3,209 \\
Country (WEIRD) & Q5 Eventual venue & degen. & degen. & 1,917 \\
Gender & Q1 Citations & degen. & $0.040$ (0.014) & 3,840 \\
Gender & Q2 Disruption & $0.046$ (0.048) & $0.036$ (0.047) &   573 \\
Gender & Q3 Novelty (embed.) & degen. & $0.043$ (0.044) & 2,267 \\
Gender & Q4 Novelty (atyp.) & $-0.022$ (0.033) & $-0.019$ (0.046) & 3,455 \\
Gender & Q5 Eventual venue & degen. & degen. & 2,301 \\
\midrule
\multicolumn{5}{l}{\emph{Panel B: revision extent regressed on group}} \\
Axis & Measure & Estimate (SE) & $p$ & \\
\midrule
Institutional prestige & Text similarity & $-0.026$ (0.015) & 0.090 & \\
Institutional prestige & Days to republication & degen. & --- & \\
Institutional prestige & Venue jump (tiers) & $-0.137$ (0.051) & 0.007 & \\
Country (WEIRD) & Text similarity & $-0.043$ (0.014) & 0.002 & \\
Country (WEIRD) & Days to republication & degen. & --- & \\
Country (WEIRD) & Venue jump (tiers) & $-0.295$ (0.052) & $<\!10^{-7}$ & \\
Gender & Text similarity & $-0.012$ (0.012) & 0.337 & \\
Gender & Days to republication & degen. & --- & \\
Gender & Venue jump (tiers) & $-0.089$ (0.047) & 0.061 & \\
\bottomrule
\end{tabular}

\end{table}

\section{Pre-analysis plan}
\label{app:pap}

The registered pre-analysis plan is available, with its timestamped registration, at \url{https://osf.io/ndkuq}.

\end{document}